\documentclass[reprint,amsmath,amssymb,aps,superscriptaddress,twocolumn,10pt]{revtex4-1}
\usepackage[colorlinks=true,allcolors=blue]{hyperref}
\usepackage{graphicx,color}
\usepackage{dcolumn}
\usepackage{subfigure}
\usepackage{bm}
\usepackage{amsmath,amsfonts,amssymb}
\usepackage{acronym}
\usepackage{enumitem}
\usepackage{array}
\usepackage{multirow}
\usepackage{CJK}
\allowdisplaybreaks
\newcommand{\be}{\begin{equation}}
\newcommand{\ee}{\end{equation}}
\newcommand{\bea}{\begin{eqnarray}}
\newcommand{\eea}{\end{eqnarray}}

\newcommand{\Msun}{{\rm M}_\odot}

\newcommand{\SOUTHCUT}{
School of Physics and Optoelectronics, South China University of Technology, Guangzhou 510641,
China}
\newcommand{\NCUa}{Department of physics, Nanchang University, Nanchang, 330031, China}
\newcommand{\NCUb}{Center for Relativistic Astrophysics and High Energy Physics, Nanchang University, Nanchang, 330031, China}

\newacro{EMRI}{extreme mass-ratio inspirals}
\newacro{MBH}{massive black hole}
\newacro{BH}{black hole}
\newacro{GR}{general relativity}
\newacro{HKBH}{hairy Kerr black hole}
\newacro{KNBH}{Kerr-Newmann black hole}
\newacro{KBH}{Kerr black hole}
\newacro{NHT}{no-hair theorem}
\newacro{DWD}{double white dwarf}
\newacro{GW}{gravitational wave}
\newacro{AK}{analytic kludge}
\newacro{NK}{numerical kludge}
\newacro{AAK}{augmented analytic kludge}
\newacro{CO}{compact object}
\newacro{PE}{parameter estimation}
\newacro{SNR}{signal-to-noise ratio}
\newacro{PN}{post newtonion}
\newacro{FIM}{Fisher information matrix}
\newacro{LSO}{last stable orbit}
\newacro{ISCO}{innermost stable circular orbit}
\newacro{BBH}{Binary Black Hole}
\newacro{BNS}{Binary Neutron Star}
\newacro{NS}{Neutron Star}
\newacro{KN}{Kerr-Newmann}
\newcommand{\beq}{\begin{equation}}
\newcommand{\eeq}{\end{equation}}
\newcommand{\beqa}{\begin{eqnarray}}
\newcommand{\eeqa}{\end{eqnarray}}

\def\lsim{\mathrel{\rlap{\lower4pt\hbox{\hskip0.5pt$\sim$}}
    \raise1pt\hbox{$<$}}}         %less than or approx. symbol
\def\gsim{\mathrel{\rlap{\lower4pt\hbox{\hskip0.5pt$\sim$}}
    \raise1pt\hbox{$>$}}}         %greater than or approx. symbol
\begin{document}
\begin{CJK*}{UTF8}{gbsn}
\title{Constraint on massive vector field with extreme-mass-ratio inspirals around a slowly rotating black hole}
\author{Tieguang Zi}
\affiliation{\NCUa}
\affiliation{\NCUb}
\author{Peng-Cheng Li}
\email{pchli2021@scut.edu.cn}
\affiliation{\SOUTHCUT}

\author{Bao-Min Gu}
\email{gubm@ncu.edu.cn}
\affiliation{\NCUa}
\affiliation{\NCUb}

\author{Fu-Wen Shu}
\email{fuwenshu@ncu.edu.cn}
\affiliation{\NCUa}
\affiliation{\NCUb}

\begin{abstract}
We study the influence of a massive vector (Proca) field on the energy fluxes from extreme-mass-ratio inspirals (EMRIs) around a slowly rotating Kerr black hole. The secondary compact
object, carrying a Proca hair, emits additional dipolar radiation that alters total energy flux
relative to general relativity (GR). These modifications induce a secular drift in the orbital evolution of circular geodesic orbits, leading to measurable dephasing in the resulting EMRIs waveforms. By evaluating
waveform mismatches between the Einstein-Proca framework and its GR counterpart, we show that the
Laser Interferometer Space Antenna (LISA) can distinguish the signatures of a light Proca field
when black hole rotation is included. Furthermore, using a Fisher information matrix analysis, we forecast LISA's capability to place stringent constraints on the Proca mass with EMRIs signal from slowly rotating Kerr black holes. For representative EMRIs configurations, we find that LISA can detect or constrain Proca masses down to $\mu_v\sim 10^{-20}$eV, with typical fractional uncertainties at the level of tens percent, depending on the black-hole spin.
\end{abstract}

\maketitle
\end{CJK*}
\section{Introduction}
Gravitational-wave (GW) detections from compact binary systems have revolutionized experimental tests of
general relativity (GR) with unprecedented precision~\cite{LIGOScientific:2025pvj,LIGOScientific:2025yae,LIGOScientific:2025obp}, particularly following the loudest event recorded during the O4 observing run of the LIGO-Virgo-KAGRA Collaboration~\cite{LIGOScientific:2025cmm,KAGRA:2025oiz}. From Solar System experiments~\cite{Will:2014kxa}
to binary black hole and neutron star mergers~\cite{LIGOScientific:2016aoc,LIGOScientific:2017vwq,KAGRA:2021vkt,LIGOScientific:2021sio,Xie:2024xex,Berti:2025hly,Yunes:2025xwp},
all available observations continue to confirm the predictions of GR with remarkable accuracy. In particular, the nature of rotating black holes (BHs), as encapsulated by the
no-hair theorem and Hawking's area law, remains consistent with the observational evidence to date. Despite
these successes, Einstein's theory faces several profound theoretical challenges, including its incompatibility with
quantum mechanics~\cite{Rovelli:1997yv,Carlip:2001wq}, the existence of spacetime singularities~\cite{Penrose:1964wq,Hawking:1970}, and the unexplained late-time acceleration of cosmic expansion~\cite{Riess:1998cb,Perlmutter:1998np,Spergel:2003cb,Planck:2015xua}.
To address these issues, numerous extensions and modifications of GR have been proposed. Observations of GWs from compact objects
provide a unique opportunity to test such theories in the strong-field, highly dynamical regime, thus offering
stringent constraints on possible deviations from GR~\cite{Will:2014kxa}.

In the coming decade, next-generation space-based GW detectors such as the Laser Interferometer Space Antenna (LISA)~\cite{LISA:2017pwj}, TianQin~\cite{TianQin:2015yph,TianQin:2020hid} and Taiji~\cite{Ruan:2018tsw} are expected to detect high signal-to-noise-ratio (SNR) signals from more massive and distant sources. Among these, extreme mass-ratio inspirals (EMRIs) systems in which a stellar-mass compact object (secondary, $m_p \sim 1$--$100 M_\odot$) orbits a massive black hole (primary, $M \sim 10^5$--$10^7 M_\odot$) are of particular interest. Due to their small mass ratio, the secondary typically completes $10^4$--$10^5$ orbital cycles before plunging, allowing EMRIs to serve as exquisite probes of the spacetime geometry surrounding massive black hole (MBH). Accurately modeling EMRI dynamics and waveforms within GR remains an ongoing challenge that has occupied decades of theoretical development~\cite{Barack:2018yvs}.
Building upon the Teukolsky formalism, modern computational methods now allow for accurate modeling of circular, inclined and generic orbits~\cite{Hughes:1999bq,Hughes:2001jr,Drasco:2005kz,Hughes:2005qb,Sago:2016xsp,Fujita:2020zxe,Hughes:2021exa,Isoyama:2021jjd,Piovano:2024yks,Drummond:2022efc,Skoupy:2023lih,Rink:2024swg,Chapman-Bird:2025xtd,Chen:2025ncm}.

Extending these analyses of EMRIs waveforms to theories beyond GR introduces additional complexity, primarily due to two key difficulties: (1) the lack of a unified perturbative framework capable of describing EMRIs orbital dynamics in generic spacetimes, and (2) the presence of additional polarization modes and couplings in modified gravity theories. Consequently, many beyond-GR EMRIs studies have relied on weak-field approximations, such as post-Newtonian expansions or quadrupole fluxes computed from geodesic motion neglecting radiation reaction~\cite{Barausse:2006vt,Yunes:2011aa,Gair:2011ym,Pani:2011xj,Sopuerta:2009iy,Canizares:2012is,
Moore:2017lxy,Zi:2022hcc,Zi:2024jla,Qiao:2024gfb,Kumar:2024our,Fu:2024cfk,Zhang:2024csc,Zhao:2024exh,
Yang:2024cnd,Gu:2024dna,Wang:2025hla,Zhao:2025sck,Xia:2025yzg,Yang:2025esa,Zare:2025aek,Zhang:2025wni}.
Over the past several years, significant progress has been made in constructing theoretical EMRIs waveform templates that incorporate diverse astrophysical effects. These include tests of GR and its alternatives~\cite{Glampedakis:2005cf,Barack:2006pq,Babak:2017tow,Speri:2024qak,Zi:2024dpi,Barsanti:2024kul}, investigations of dark compact object candidates~\cite{Maggio:2021ans,Maggio:2021uge,Cardoso:2022fbq,Zi:2024itp}, studies of environmental effects in galactic nuclei~\cite{Barausse:2014tra,Hannuksela:2018izj,Cardoso:2019upw,Toubiana:2020drf,DeLuca:2022xlz,Cole:2022yzw,Dyson:2025dlj,Becker:2024ibd,Pan:2021ksp,Zwick:2025wkt,Kejriwal:2023djc,Rahman:2025mip,Copparoni:2025vty,Kejriwal:2025jao}, and analyses of accretion processes between massive black holes and their companions~\cite{Kocsis:2011dr,Zwick:2021dlg,Caputo:2020irr,Speri:2022upm,Pan:2021oob,Li:2025zgo,Lyu:2024gnk,HegadeKR:2025dur,Lam:2025fzi}.

From the perspective of fundamental physics, EMRIs also provide a promising avenue for detecting new degrees of freedom beyond GR. Motivated by early work on model-independent searches for scalar charges~\cite{Maselli:2020zgv}, recent studies have extended these efforts to probe scalar~\cite{Maselli:2020zgv,Maselli:2021men,Barsanti:2022ana,Barsanti:2022vvl,DellaRocca:2024pnm,Barsanti:2024kul,Speri:2024qak,Jiang:2021htl,Guo:2022euk,Zhang:2022rfr}, vector~\cite{Zhang:2022hbt,Zi:2022hcc,Liang:2022gdk,Zhang:2024ogc,Zi:2024lmt} and tensor~\cite{Cardoso:2018zhm} fields. These developments have paved the way for computing additional radiation channels associated with new fundamental fields, such as scalar or vector degrees of freedom in modified gravity scenarios. In addition to their role in modified-gravity scenarios, massive test fields are also well-motivated from the perspective of particle physics and cosmology. Ultra-light bosonic fields arise naturally in string-inspired axiverse models, hidden
U(1) extensions, and various constructions of ultralight dark matter \cite{Hui:2021tkt}. Such fields can form long-lived condensates around astrophysical BHs, or leave imprints on GW signals from compact binaries. Therefore, EMRIs observed by LISA provide not only a test of dynamical modifications to GR, but also a novel probe of ultra-light  dark matter over a mass range that is difficult to access by laboratory or cosmological observations \cite{LISA:2022kgy}.

In this paper, we explore the detectability of ultra-light vector fields with LISA through their impact on circular EMRIs around slowly rotating Kerr BHs. Previous works investigated the constraint on the massive vector hair carried by secondary object on different orbital types of EMRIs around Schwarzschild BH~\cite{Zi:2024lmt,Zi:2025qos}. However, since astrophysical BHs are generically expected to possess nonzero spin, it is essential to extend previous Schwarzschild-based studies to the Kerr case. We compute the gravitational and Proca fluxes using perturbation theory, evolve the adiabatic inspiral trajectories, and construct the corresponding quadrupolar waveforms. Using the Fisher information matrix formalism, we estimate LISA's ability to measure the Proca field mass and assess how this sensitivity depends on the spin of the central BH.

Although the Proca equation on a generic Kerr background is known to be separable \cite{Frolov:2018ezx}, the resulting system of coupled radial and angular equations is significantly more involved than in the scalar case. In particular, the frequency-domain formulation involves coupled polar and axial sectors as well as higher-order spin interactions, and a practical scheme to compute gravitational and Proca energy fluxes for generic Kerr spin has not yet been established for EMRI applications. For this reason, the slowly rotating limit provides a technically tractable and well-controlled framework in which the essential Proca-induced modifications to the inspiral dynamics can be computed reliably~\cite{Pani:2012bp}. Our analysis therefore constitutes a first step toward understanding massive-vector effects in EMRIs, and can serve as a baseline for future extensions to moderate and high spins.

The paper is organized as follows. In Sec.~\ref{method}, we present the theoretical framework of EMRIs in the family of Einstein-Proca gravity, describing the relevant perturbation equations and their effects on EMRIs flux. Section~\ref{result} discusses our main results on LISA's detection prospects and parameter estimation. Finally, Sec.~\ref{conclusion} summarizes our findings and outlines future directions.
\section{Method}\label{method}
\subsection{Setup}
We begin to consider the action as follows
\begin{equation}\label{action}
S(g,A,\Psi) = S_0(g,A)+ \alpha S_c(g, \Psi) + S_m(g,A,\Psi)\;,
\end{equation}
where $A$ represents the Proca field with a mass $\mu_v$.
We here introduce a dimensionless parameter $\mu\equiv\mu_v M$, with $\mu_v$  defined by electronvolts in geometric units $(G=c=1)$~\cite{Brito:2015oca,Barsanti:2022vvl,Zi:2025qos},
\begin{align}
\mu_v[{\rm eV}] &\sim \left(\frac{\mu_v M}{0.75}\right) \left(\frac{10^6 M_\odot}{M}\right) 10^{-16} \rm eV
\nonumber \\   & = \left(\frac{\mu}{0.75}\right) \left(\frac{10^6 M_\odot}{M}\right) 10^{-16} \rm eV\;.
\end{align}
Here the quantity $S_0(g,A)$ is the action describing spacetime geometry, $\alpha S_c(g, \Psi)$ encodes the nonminimal coupling between the metric tensor $g$ and matter field $\Psi$. The parameter $\alpha$ is a constant with the dimension $[\alpha]=(\rm mass)^n$, $n$ is a positive integer.  $S_m(g,A,\Psi)$ is the action depending on the matter field $\Psi$.
In this work, we assume that a Proca field is minimally coupled to the metric tensor $g$, which is determined by the following action
\begin{eqnarray}\label{Lagrangian}
S_0(g,A)&= &\int d^4x \sqrt{-g}
\left[\frac{1}{16\pi } R - \mathcal{L}\right]
\end{eqnarray}
and
\begin{equation}
\mathcal{L} = \frac{1}{4}F_{\mu\nu}F^{\mu\nu}+ \frac{\mu^2}{2}A_{\mu}A^\mu,
\end{equation}
where $R$ is the Ricci scalar regarding to metric $g$, $F_{\mu\nu}=\nabla_\mu A_{\nu}- \nabla_\nu A_{\mu}$ is the strength of Proca field with a mass $\mu_v$.
From the action~\eqref{action}, varying with respect to the metric and the Proca field, we obtain the equation of motion for two fields
\begin{eqnarray}
G^{\mu\nu}=R^{\mu\nu} -\frac{1}{2} g^{\mu\nu} R&=& \mathcal{T}^{\mu\nu}_{\rm Proca} \nonumber  +  \mathcal{T}_p^{\mu\nu} + \alpha \mathcal{T}^{\mu\nu}_c  \label{eq:metric}\;, \\
\nabla_\rho F^{\rho\mu}-\mu_v^2 A^\mu + \frac{8\pi\alpha}{\sqrt{-g}} \frac{\delta S_c}{\delta A^\mu}& =& 8\pi J^\mu \;, \label{eq:vector}
\end{eqnarray}
where $J^\mu$ is the current density for vector field
\begin{equation}
J^\mu = q m_p \int d\tau u^\mu \frac{\delta^{4} [x-y(\tau)]}{\sqrt{-g}}\;.
\end{equation}
with a vector charge $q$ carried by secondary object with mass $m_p$, the quantity $u^\mu=dx^\mu/d\tau$
is the four-velocity of point particle and $y(\tau)$ describes its the worldline parametrized by the proper time $\tau$. Here the symbol $\delta^{4}[\cdot]$ is the four-dimensional delta distribution function.

We begin to introduce an approximation scheme to simplify the two equations in~\eqref{eq:vector}.
The coupling constant $\alpha$ and the mass ratio $\eta$ are related through
\begin{equation}
\frac{\alpha}{M^n} = \frac{\alpha}{m_p^n} \eta^n,
\end{equation}
where $\alpha/m_p^n$ is a quantity of order $\mathcal{O}(1)$ or smaller.
This follows from the fact that current astrophysical observations have not revealed any deviations from the Kerr hypothesis in either the weak and strong field regimes~\cite{Will:2014kxa,LIGOScientific:2025cmm,KAGRA:2025oiz}.
In the special case $\alpha = 0$, the no-hair theorem still holds within modified gravity, implying that the primary object can be described by the Kerr metric.
Therefore, the exterior spacetime can be interpreted as a small perturbative correction to the Kerr black hole, characterized by a dimensionless parameter
\begin{equation}
\xi = \frac{\alpha}{M^n} = \eta^n \frac{\alpha}{m_p^n}.
\end{equation}
Given that $\alpha/m_p^n \leq 1$ and the mass ratio $\eta \ll 1$, the primary black hole in the theories described by~\eqref{action} can be well approximated by a Kerr spacetime.

Toward the following step, we introduce the perturbation method using mass-ratio $\eta$ as a bookmark,
in which the tensor and vector fields are expanded as
\begin{equation}
g_{\mu\nu}=g_{\mu\nu}^{(0)}+\eta h_{\mu\nu}^{(1)}+ \cdots \;,  \quad
A_\mu=A_\mu^{(0)}+\eta A_\mu^{(1)}+\cdots\;,.
\end{equation}
Here the $A_\mu^{(0)}$ is a constant background field due to the constraint of no-hair theorem, which can be set to zero. $g_{\mu\nu}^{(0)}$ is  tensor field, determining by Kerr metric.
At the first perturbation of mass-ratio of $\mathcal{O}(\eta)$, the perturbations  of two fields are excited by secondary. Using the skeletonized method~\cite{Ramazanoglu:2016kul,Maselli:2020zgv,Maselli:2021men,Barsanti:2022vvl}, the matter action $S_m$ can be substituted by an action of the point particle $S_p$.
The massive, Proca-haired secondary object have a such action
\begin{equation}
S_p = -\int m(A_\mu) \sqrt{g_{\mu\nu} \frac{dx^\mu}{d\lambda}\frac{dx^\nu}{d\lambda}} d\lambda\;.
\end{equation}
Following the method in Ref.~\cite{Maselli:2021men}, the mass function $m(A_\mu)$ depends on the
vector field in a reference frame $\tilde{x}$ centered on secondary.
The vector perturbation can be written as follows:
\begin{equation}
A_\mu^{(1)} \approx \frac{m_p}{\tilde{r}}q e^{-\mu_v \tilde{r}} + \mathcal{O}\left(\frac{m_p^2}{\tilde{r}^2} e^{-\mu_v \tilde{r}}\right)\;,
\end{equation}

In this framework, $\delta{S_c}/\delta A_\mu$ and the stress-energy tensor of vector filed $\mathcal{T}_{\rm Proca}^{\mu\nu}$ are quadratic terms of mass-ratio $\mathcal{O}(\eta^2)$. The leading contribution at first order $\mathcal{O}(\eta)$ should result from the  stress-energy tensor of secondary. Therefore, the equation should be reduced as
\begin{eqnarray}
G_{\mu\nu}^{(1)} &=& 8\pi m_p  \int d\tau u_\mu u_\nu \frac{\delta^{4} [x-y(\tau)]}{\sqrt{-g}}
\label{eq:einstein} ~\;,  \nonumber \\
\nabla^{\rho} F_{\rho\mu}^{(1)} - \mu_v^2 A_{\mu}^{(1)}& =& 4\pi q m_p  \int d\tau u_\mu \frac{\delta^{4} [x-y(\tau)]}{\sqrt{-g}}  ~\;,\label{eq:vector}
\end{eqnarray}
where the superscript denotes the tensor or vector quantity as a function of perturbation field.
The two equations can be solved with perturbative theory~\cite{Teukolsky:1973ha}, where the gravitational equation can be studied extensively in the Teukolsky formalism and their fluxes from EMRIs can be computed using \texttt{C++} or \texttt{Mathematica}  packages in \texttt{Black Hole Perturbation Toolkit} \cite{BHPToolkit}.
In the following section we lay emphasis on the massive vector radiation for EMRIs around a slow-rotating Kerr BH and compute its vector fluxes to evolve orbital parameters adiabatically.
\subsection{Massive vector perturbation around slowly spinning Kerr black hole}
Following the vector-field decomposition scheme developed in Refs.~\cite{Gerlach:1980tx,Pani:2012bp}, the Proca equation can be reduced to four coupled ordinary differential equations (ODEs) through expansion in vector spherical harmonics within the Regge-Wheeler-Zerilli gauge~\cite{Rosa:2011my}.
We first consider the perturbations of a slowly rotating black hole, whose most general stationary and axisymmetric spacetime in coordinates $(t,r,\theta,\phi)$ can be given by
\begin{equation}
ds^2=H^2dt^2+Q^2dr^2+r^2K^2[d\theta^2+\sin^2\theta(d\phi-Ldt)^2]\;, \label{general:spacetime}
\end{equation}
where the functions $(K,H,Q,L)$ depend on coordinates $(r,\theta)$.
Up to first order of rotation, the background \eqref{general:spacetime} can be  written as  ~\cite{Pani:2012bp}
\begin{eqnarray}
ds^2=&-&F(r)dt^2+B(r)^{-1}dr^2+2\varpi(r)\sin^2\theta d\phi dt
 \nonumber \\ &-&r^2(\sin^2\theta d\theta^2+d\phi^2)   \;, \label{first:spacetime}
\end{eqnarray}
where $F(r)=B(r)=1-2M/r$ and $\varpi(r)=2M^2a/r$ for a slowly rotating BH, $M$ is mass of BH
and $a=J/M$ is the dimensionless parameter of angular momentum of rotating BH.
Under this background, the electronmagnetic potential and the current density of four-vector can be decomposed into
\begin{equation}
A_\mu(t,r,\theta,\phi) = \sum_{\ell m}
\begin{bmatrix}
0 \\ 0 \\  u_{4}^\ell \mathbf{S}_b^{\ell m}/\Lambda
\end{bmatrix} +  \sum_{\ell m}
\begin{bmatrix}
 u_{1}^\ell Y^\ell/r \\ u_{2}^\ell Y^{\ell m}/(rF)  \\  u_{3}^\ell \mathbf{Y}^{\ell m}_b/\Lambda\;,
\end{bmatrix}
\end{equation}
\begin{equation}
J_\mu(t,r,\theta,\phi) = \sum_{\ell m}
\begin{bmatrix}
 s_{(2)}Y^{\ell m}/(rF) \\  s_{(3)} \mathbf{Y}^{\ell m}_b/\sqrt{\Lambda} \\  s_{(4)} \mathbf{S}_b^{\ell m}/\sqrt{\Lambda}
\end{bmatrix} \;,
\end{equation}
where $\Lambda=\ell(\ell+1)$, the quantities $s_{(4)}$ can be obtained with the perturbation information of secondary body on the geodesic of a slowly rotating BH
and the vector spherical harmonics can be defined as
\begin{eqnarray}
\mathbf{Y}_b^\ell & \equiv & (\partial_\theta ,\partial_\phi )Y^\ell \;, \\
\mathbf{S}_b^\ell & \equiv & \left(\frac{1}{\sin\theta}\partial_\phi ,-\sin\theta\partial_\theta \right)Y^\ell \;. \label{eq:vector:expand}
\end{eqnarray}
Here $Y^{\ell m}\equiv Y^{\ell m}(\theta,\phi)$ is the spherical harmonics in a spherical spacetime, the functions $(u^\ell_{1,2,3})$ describe the polar sector and $(u^\ell_{4})$ determines the axial sector.
Using the  expansion of vector potential~\eqref{eq:vector:expand}, the Proca equation can be reduced to
four ODEs regarding to $u^\ell_{i}$ $(i=1,2,3,4)$, then one can obtain the system of equations for the polar
and axial sector with Lorenz identity $(\nabla_\mu A^\mu=0)$.
For the polar sector, the ODEs of $u^\ell_{2,3}$ can be finally simplified as Eqs.(63-65) in Ref.~\cite{Pani:2012bp}
\begin{align}
&\mathcal{D}_2 u_{2}^\ell- \frac{2F}{r^2}\left(1-\frac{3M}{r}\right)(u_{2}^\ell-u_{3}^\ell) \nonumber\\
&=\frac{2amM^2}{\Lambda r^5 \omega}\left[\Lambda (2r^2\omega^2+3F^2)u_{2}^\ell
\right.\nonumber\\ &\left.
+3F(r\Lambda F u_{2}'^\ell - (r^2\omega^2 + \Lambda F) u_{3}^\ell ) \right]
 + 4\pi s_{(2)}F \;, \label{eq:u2} \\
&\mathcal{D}_2 u_{3}^\ell + \frac{2F}{r^2} u^{\ell}_2 = \frac{2amM^2}{r^5\omega}\left[2r^2\omega^2 u^{\ell}_3 + 3 rF^2u'^{\ell}_3 \right.\nonumber\\ &\left.  -
3(\Lambda+r^2\mu^2)F u^{\ell}_2
\right] + 4\pi s_{(3)}F  \;.\label{eq:u3}
\end{align}
The axial sector is
\begin{align}
\mathcal{D}_2 u_{4}^\ell - \frac{4aMm\omega}{r^3} u^{\ell}_4 &= 4\pi s_{(2)}F \;,  \label{eq:u4}
\end{align}
where $\mathcal{Q}_\ell=\sqrt{(\ell^2-m^2)/(4\ell^2-1)}$ and $\psi^\ell=(\Lambda+r^2\mu^2 ) u^{\ell}_2- (r-2M) u'^{\ell}_3$.
The superscript $\ell$ in functions $u^{\ell}_{2,3,4}$ denotes the perturbation of different order  in the vector spherical harmonics, where the perturbation for $\ell$ is only coupled to that for $\ell+1$ mode~\cite{Pani:2012bp}.
The operator is
\begin{equation}
\mathcal{D}_2  = \frac{d^2}{dr_\ast^2} + \omega -F\left[\mu^2+\frac{\ell(\ell+1)}{r^2} \right]\;,
\end{equation}
where the derivative $d/dr_\ast$ is taken with respect to the tortoise coordinate $r_\ast = r+2\ln\left[r/(2M)-1\right]$.

To obtain the homogeneous solutions of Eqs.~\eqref{eq:u2}--\eqref{eq:u4}, we need to impose the
ingoing and outgoing boundaries as follows
\begin{eqnarray}
u_{2,3,4}^{\ell,H} &=&\sum_{n=0}^{n_H}e^{-i \omega r_\ast(r)}a_n(r-r_h)^n \;,\\
u_{2,3,4}^{\ell,\infty} &=&\sum_{n=0}^{n_\infty}e^{i \sqrt{\omega^2-\mu^2} r_\ast(r)}r^{\frac{iM\mu^2}{\sqrt{\omega^2-\mu^2}}}\frac{b_n}{r^n},
\end{eqnarray}
which can yield the ingoing solution near the horizon $(r_h)$ and outgoing solution
at radial infinity.  The coefficients $(a_n,b_n)$ are determined by soloving homogeneous
equations~\eqref{eq:u2}--\eqref{eq:u4} order by order in $r-r_h$ and $1/r$ and leading terms are set to
$a_0=b_0=1$.
These coefficients of boundary conditions, together with the higher-order expansion, ensure numerical stability and high accuracy in the integration setting $n_H=6$ and $n_\infty=8$.
Using the Green's function method, the homogeneous solutions of the Proca field equation \eqref{eq:vector} are obtained by imposing two boundary conditions, which are integrated over source term to obtain the non-homogeneous solutions.
\subsection{Source term of vector perturbation induced by secondary}
In this subsection, we investigate the dynamics of EMRIs, focusing on the vector perturbations generated by a point particle orbiting a MBH.
Assuming that the inspiraling particle follows a circular geodesic on the equatorial plane, i.e.,
$(r=r_p, \theta=\pi/2, \phi=\Omega_p t)$ with orbital frequency $\Omega_p=\sqrt{M}/(r_p^{3/2}+a\sqrt{M})$.
The current density for a massive vector field can be given by
\begin{equation}
J^\mu=q\frac{u^\mu}{u^t r_p^2 \sin\theta_p}\delta(r-r_p)\delta(\theta-\theta_p)\delta(\phi-\phi_p),
\end{equation}
where four vector of secondary is
\begin{equation}
u^{\mu}=\left(E_p/F,0,0,L_p/r_p^2\right)
\end{equation}
with orbital energy and angular momentum
\begin{eqnarray}
\frac{E_p}{m_p} = \frac{1-2v_p^2+av_p^3}{\sqrt{1-3v_p^2+2av_p^3}}\;,\\
\frac{L_p}{m_p} = \pm v_p \frac{1-2v_p^3+a^2v_p^4}{\sqrt{1-3v_p^2+2av_p^3}}\;.
\end{eqnarray}
Here $v_p\equiv \sqrt{M/r_p}$ is the orbital velocity of point particle, and
the plus and minus symbol denote prograde and retrograde orbits, we only focus on the prograde orbits in this work.
The source terms for vector perturbed equations~\eqref{eq:u2}--\eqref{eq:u4} are given by
\begin{equation}\label{source:term:coeffs}
\begin{split}
s_{(2)}&=0,\\
s_{(3)}&=-q\frac{\partial_\phi Y^{lm}(\pi/2,0)}{r_p^{3/2}}\delta(r-r_p)\delta(\omega-\omega_m),\\
s_{(4)}&=-q \frac{\partial_\theta Y^{lm}(\pi/2,0)}{r_p^{3/2}}\delta(r-r_p)\delta(\omega-\omega_m),
\end{split}
\end{equation}
where $\omega_m = m \Omega_p$ is the fundamental frequency for the circular and equatorial geodesic of a slowly spinning Kerr BH.

\textbf{Axial \& Polar inhomogeneous solutions:}
Using the source term~\eqref{source:term:coeffs}, one can obtain the inhomogeneous solutions using the purely ingoing and outgoing solutions as~\cite{Zi:2024lmt}.
For the axial section, the general solution
\begin{equation}
u_{(4)}=u_{4}^{H}\int_{r}^{\infty}d r_\ast~ \frac{4\pi f s_{(4)}}{W}u_{4}^{\infty}+u_{4}^{\infty}\int_{-\infty}^{r}d r_\ast~ \frac{4\pi f s_{(4)}}{W}u_{4}^{H} \, ,
\end{equation}
with the wronskian
\begin{equation}
W=\frac{du_{4}^{\infty}}{d r_\ast}u_{4}^{H}- \frac{du_{4}^{H}}{d r_\ast}u_{4}^{\infty}.
\end{equation}
Here we omit the superscript $\ell$ of $u_{4}^{\ell, H,\infty}$ for simplicity in the following discussion.
So the inhomogeneous solutions for the axial mode near the horizon and at infinity
\begin{align}
\begin{split}
\lim_{r_\ast\to -\infty}u_{(4)}&=e^{-i \omega r_\ast(r)}\int_{-\infty}^{\infty}d r_\ast~\frac{4\pi f s_{(4)}}{W} u_{4}^{H}\\
&\equiv\mathcal{A}^{4-}_{lm}e^{-i \omega r_\ast(r)}, \\
\lim_{r_\ast\to +\infty}u_{(4)}&= e^{i \sqrt{\omega^2-\mu^2} r_\ast(r)}r^{\frac{iM\mu^2}{\sqrt{\omega^2-\mu^2}}}\int_{-\infty}^{\infty}d r_\ast~\frac{4\pi f s_{(4)}}{W} u_{4}^{\infty} \\
&\equiv\mathcal{A}^{4+}_{lm}e^{i \sqrt{\omega_m^2-\mu^2} r_\ast(r)}r^{\frac{iM\mu^2}{\sqrt{\omega^2-\mu^2}}}.
\end{split}
\end{align}

For the polar mode, the functions $u_{(2)}$ and $u_{(3)}$ in Eqs.~\eqref{eq:u2} and \eqref{eq:u3} are coupled each other, which can be systematized into a coupled system in the following matrix form
\begin{align}
\label{eq:polar:u2u3:master}
\left( \mathcal{D}_2 + \omega^2
+ \left[\begin{array}{cc} \alpha_{lm} & \beta_{lm} \\ \gamma_{lm} & \sigma_{lm} \end{array} \right] \right) \left[\begin{array}{c} u_{2} \\ u_{3} \end{array} \right]
&+
\left[\begin{array}{cc}  \varsigma & 0 \\  \varsigma  & 0 \end{array} \right]
\left[\begin{array}{c} u'_{2} \\ u'_{3} \end{array} \right]
\nonumber\\ &= \left[\begin{array}{c} S_{lm} \\ Z_{lm} \end{array} \right],
\end{align}
where
\begin{equation}
\begin{split}
\alpha_{lm}&=- \frac{2F}{r^2}\left(1-\frac{3M}{r}\right) - \frac{2a^2mM^2}{r^5\omega} \left(2r^2\omega^2+3F^2\right) \;,\\
\beta_{lm}&= \frac{2F}{r^2}\left(1-\frac{3M}{r}\right) + \frac{6Fa^2mM^2}{r^5 \Lambda\omega} \left(r^2\omega^2+\Lambda F^2\right) ,\\
\gamma_{lm}&=\frac{2F\Lambda}{r^2} + \frac{6Fa^2mM^2}{r^5 \Lambda\omega} \left(r^2\mu^2+\Lambda \right)  \;,\\
\sigma_{lm}&=- \frac{4amM^2\omega}{r^3} \;,\\
\varsigma &=  \frac{6amM^2 F}{r\omega} \;,
\end{split}
\end{equation}
they can be solved using Green's function method to obtain the ingoing and outgoing homogeneous solutions.
The source term in Eq.~\eqref{eq:polar:u2u3:master} can be recast as
\begin{eqnarray}
\label{eq:source}
\left[\begin{array}{c} S_{lm}(r) \\ Z_{lm}(r) \end{array} \right] &=& \left[\begin{array}{c} B_{lm} \\ D_{lm} \end{array} \right]\delta(r-r_0) \nonumber \\ &+& \left[\begin{array}{c} F_{lm} \\ H_{lm} \end{array} \right] \frac{d}{dr}\delta(r-r_0) ,
\end{eqnarray}
where  $B_{lm}$, $D_{lm}$, $F_{lm}$, and $H_{lm}$ are determined by the circular geodesic of the point particle.
The general solution can be obtained by matching to homogeneous solution at the horizon and infinity
\begin{align}
\label{eq:inhomo}
&\left[\begin{array}{c} u_{(2)} \\ u_{(3)} \end{array} \right] = \left( \mathcal{A}^{2+}_{lm} \left[ \begin{array}{c} u_{2}^{0+} \\ u_{3}^{0+} \end{array} \right] + \mathcal{A}^{3+}_{lm} \left[ \begin{array}{c} u_{2}^{1+} \\ u_{3}^{1+} \end{array} \right]  \right) \Theta(r-r_p) \notag
\\&\qquad\;\;\; +\left( \mathcal{A}^{2-}_{lm} \left[ \begin{array}{c} u_{2}^{0-} \\ u_{3}^{0-} \end{array} \right] + \mathcal{A}^{3-}_{lm} \left[ \begin{array}{c} u_{2}^{1-} \\ u_{3}^{1-} \end{array} \right]  \right) \Theta(r_p-r) ,
\end{align}
where $\Theta$ is the Heaviside step function and the superscript $\pm$ of $u_{2,3}^\pm$ denotes the
outgoing and ingoing homogeneous solutions at the horizon and at infinity, respectively.
For the spatial infinity $r_*=+\infty$, two outgoing homogeneous solutions when $r\ll |\omega_m|^{-1}$ can be written as
\begin{eqnarray}
\label{eq:out}
\left[ \begin{array}{c} u_{2}^{0+} \\ u_{3}^{0+} \end{array} \right] \simeq
e^{i M\sqrt{\omega^2-\mu^2} r_\ast(r)}r^{\frac{iM\mu^2}{\sqrt{\omega^2-\mu^2}}} \left[ \begin{array}{c} 1 \\ 0 \end{array} \right] ,\nonumber \\
 \;\;\;\;\; \left[ \begin{array}{c} u_{2}^{1+} \\ u_{3}^{1+} \end{array} \right] \simeq e^{i \sqrt{\omega^2-\mu^2} r_\ast(r)}r^{\frac{iM\mu^2}{\sqrt{\omega^2-\mu^2}}} \left[ \begin{array}{c} 0 \\ 1 \end{array} \right] ,
\end{eqnarray}
and for the horizon $r_*=-\infty$,  two linearly independent homogeneous solutions when $r-r_+ \ll M$ can
be given by
\begin{eqnarray}
\label{eq:down}
\left[ \begin{array}{c} u_{2}^{0-} \\ u_{3}^{0-} \end{array} \right] \simeq e^{-i\omega r_*} \left[ \begin{array}{c} 1 \\ 0 \end{array} \right] , \;\;\;\;\; \left[ \begin{array}{c} u_{2}^{1-} \\ u_{3}^{1-} \end{array} \right] \simeq e^{-i\omega r_*} \left[ \begin{array}{c} 0 \\ 1 \end{array} \right] .
\end{eqnarray}
Because the sets of ingoing and outgoing homogeneous solutions are not unique, one can construct an alternative basis by taking appropriate linear combinations of Eqs. \eqref{eq:out}--\eqref{eq:down}. This reformulated basis is particularly convenient for evaluating the inhomogeneous solutions. The homogeneous modes are expanded as power series near the respective boundaries, either
$(r\rightarrow \infty$ or $r \rightarrow r_h)$, allowing one to determine the boundary values required for integrating Eq.~\eqref{eq:polar:u2u3:master}. With these boundary conditions specified, the global homogeneous solutions can then be obtained numerically. Detailed formulations of the boundary expansions are provided in Ref.~\cite{Zhu:2018tzi}.

The coefficients $\mathcal{A}^{2 \pm}_{lm}$ and $\mathcal{A}^{3\pm}_{lm}$
are determined by imposing that the Dirac delta functions and their derivatives vanish when Eq.~\eqref{eq:inhomo} is substituted into Eq.~\eqref{eq:polar:u2u3:master}. Recognizing that the Dirac delta function represents the derivative of the Heaviside step function, this requirement is equivalent to applying the method of variation of parameters, wherein the Green's function is integrated against the source terms of the governing differential equation. The resulting relations form a linear system, expressed in terms of the Wronskian matrix, which self-consistently determines the normalization coefficients as follows
\begin{align}
&\left[\begin{array}{cccc} u_{2}^{0+} &  u_{2}^{1+}  & u_{2}^{0-} & u_{2}^{1-}
\\ u_{3}^{0+} &  u_{3}^{1+}  & u_{3}^{0-} & u_{3}^{1-}
\\ \mathcal{D}_2 u_{2}^{0+} & \mathcal{D}_2 u_{2}^{1+} & \mathcal{D}_2 u_{2}^{0-} & \mathcal{D}_2 u_{2}^{1-}
\\ \mathcal{D}_2 u_{3}^{0+} & \mathcal{D}_2 u_{3}^{1+} & \mathcal{D}_2 u_{3}^{0-} &\mathcal{D}_2 u_{3}^{1-}  \end{array} \right]_{r_0}
\cdot
\left[\begin{array}{c} \mathcal{A}^{2 +}_{lm} \\ \mathcal{A}^{3 +}_{lm}  \\ -\mathcal{A}^{2 -}_{lm}  \\ -\mathcal{A}^{3 -}_{lm} \end{array} \right] \notag
\\&\qquad\qquad\;\;\; = \frac{1}{r_p^3 F_p^2} \left[\begin{array}{c} r_p^3 F_{lm} \\ r_p^3 H_{lm} \\ r_p^3 F_p B_{lm}+2Mr_p F_{lm} \\ r_p^3 F_p D_{lm}+2M r_p H_{lm} \end{array} \right] ,
\label{eq:wron}
\end{align}
where all functions related to $r$ can be computed at $r=r_p$, including the metric function $F_p$.
\subsection{Vector fluxes and adiabatic evolution}
In the adiabatic approximation, assuming that the secondary object inspirals along a circular geodesic confined to the equatorial plane, the orbital evolution can be consistently treated within that framework. For the inspiraling secondary around MBH, the slow change in the orbital radius is governed by the energy fluxes radiated from EMRIs system. The total flux receives the contributions from both the Proca field and the gravitational sector, which together determine the dissipative dynamics of the inspiral.

To evaluate the vector-field contribution, we begin by computing the flux of massive vector field derived from Isaacson's stress-energy tensor~\cite{Isaacson:1968zza},
\begin{equation}
4\pi T_{\mu\nu}= g^{\alpha\beta}F_{\mu \alpha}F_{\nu \beta}+\mu^2 A_{\mu}A_{\nu}-g_{\mu\nu}\mathcal{L}\;.
\end{equation}
In the presence of a stationary and axisymmetric background, such as a slowly rotating Kerr BH, two Killing vectors, \(\xi^\alpha_{(t)}\) and \(\xi^\alpha_{(\phi)}\),
encode time-translation and rotational symmetries.
The associated energy flux through a surface \(\Sigma\) of constant radius \(r\) is defined as
\be
d E = -\int_\Sigma T^\mu_{~\nu}~ _{(t)}\xi^\nu d\Sigma_\mu
\ee
where $d\Sigma_\mu$ denotes the outward-oriented surface element on $\Sigma$.
Since the present of spherical symmetry for a slowly rotating Kerr black hole, the energy flux takes the compact form~\cite{Martel:2003jj}
\begin{equation} \label{dEdt:Formular}
\frac{dE^{\mathcal{P}}}{dt}
= -\epsilon\, r^2 F \int d\Omega \, T_{tr},
\end{equation}
where \( F = 1 - 2M/r \), and the parameter \(\epsilon\) labels the evaluation surface:
\(\epsilon = +1\) corresponds to the flux at spatial infinity, whereas \(\epsilon = -1\) refers to that near the horizon.
Using Eq.~\eqref{dEdt:Formular}, the asymptotic (outgoing) energy flux carried by the Proca field can be decomposed into even- and odd-parity components.
For the odd sector, one finds
\begin{equation}\label{eq:Edot:P:I:odd:new}
\left\langle \frac{dE^{\mathcal{P}}}{dt}\right\rangle_{\mathrm{odd}}^{\infty}
= \sum_{l=1}^{\infty}\sum_{m=1}^{l}
\frac{\omega_m \sqrt{\omega_m^2 - \mu^2}\,
|\mathcal{A}^{4+}_{lm}|^2}
{2\pi\, l(l+1)}\,,
\end{equation}
while the even sector yields
\begin{align}\label{eq:Edot:P:I:even:new}
\left\langle \frac{dE^{\mathcal{P}}}{dt}\right\rangle_{\mathrm{even}}^{\infty}
&= \sum_{l=1}^{\infty}\sum_{m=1}^{l}
\frac{\omega_m \sqrt{\omega_m^2 - \mu^2}\,
|\mathcal{A}^{3+}_{lm}|^2}
{2\pi\, l(l+1)} \notag\\
&\quad +
\frac{\mu^2}{2\pi}
\frac{\sqrt{\omega_m^2 - \mu^2}}{\omega_m}
|\mathcal{A}^{2+}_{lm}|^2.
\end{align}
Analogously, the near-horizon (ingoing) energy fluxes take the form
\begin{equation}\label{eq:Edot:P:H:odd:new}
\left\langle \frac{dE^{\mathcal{P}}}{dt}\right\rangle_{\mathrm{odd}}^{H}
= \sum_{l=1}^{\infty}\sum_{m=1}^{l}
\frac{\omega_m^2\, |\mathcal{A}^{4-}_{lm}|^2}
{2\pi\, l(l+1)}\,,
\end{equation}
and
\begin{align}\label{eq:Edot:P:H:even:new}
\left\langle \frac{dE^{\mathcal{P}}}{dt}\right\rangle_{\mathrm{even}}^{H}
&= \sum_{l=1}^{\infty}\sum_{m=1}^{l}
\frac{\omega_m^2\, |\mathcal{A}^{3-}_{lm}|^2}
{2\pi\, l(l+1)} \notag\\
&\quad + \frac{l(l+1)+4\mu^2}{8\pi}
|\mathcal{A}^{2-}_{lm}|^2 \notag\\
&\quad - \frac{i\omega_m}{4\pi}
\big(\mathcal{A}^{2-}_{lm} \mathcal{A}^{3-*}_{lm}
- \mathcal{A}^{3-}_{lm} \mathcal{A}^{2-*}_{lm}\big).
\end{align}
The angle brackets $\langle \,\rangle$ denote a time average, which will be  omitted hereafter for brevity.
Combining the asymptotic and near-horizon contributions from Eqs.~\eqref{eq:Edot:P:I:odd:new}--\eqref{eq:Edot:P:H:even:new},
the total energy flux carried by the Proca field can be expressed as
\begin{equation}
\dot{E}^{\mathcal{P}}
= \dot{E}_{\mathrm{odd}}^{\mathcal{P},H}
+ \dot{E}_{\mathrm{even}}^{\mathcal{P},H}
+ \dot{E}_{\mathrm{odd}}^{\mathcal{P},\infty}
+ \dot{E}_{\mathrm{even}}^{\mathcal{P},\infty},
\end{equation}
where the overdot denotes differentiation with respect to coordinate time. These expressions encapsulate the energy fluxes radiated by the massive vector field away from the system and into the horizon of BH, thereby governing the dissipative evolution of EMRI dynamics.

In the subsequent analysis, we turn to the gravitational energy flux in a slowly spinning Kerr spacetime, which can be computed from the Teukolsky formalism.
For practical computations, the \texttt{Black Hole Perturbation Toolkit}~\cite{BHPToolkit} provides numerical implementations of these equations, enabling the evaluation of the total gravitational flux, denoted by $\dot{E}^{\mathcal{G}}$.

The total energy flux emitted by EMRIs system thus receives contributions from both the gravitational and Proca sectors,
\begin{equation} \label{flux:total}
\dot{E} = \dot{E}^{\mathcal{G}} + \dot{E}^{\mathcal{P}},
\end{equation}
and this total flux governs the secular evolution of the orbiting particle on circular geodesics.

Within the adiabatic approximation, the slow inspiral of the secondary object is driven by the loss of orbital energy through both gravitational and Proca-field radiation. The corresponding evolution equations for the orbital radius and phase are
\begin{equation}\label{equation:evolution:insp}
\frac{dr_p}{dt} = -\,\dot{E}\left(\frac{dE_p}{dr_p}\right)^{-1},
\qquad
\frac{d\Phi}{dt} = \Omega_p,
\end{equation}
where $\Phi$ is the cumulative phase of GW and $\Omega_p$ is the orbital angular frequency.
By numerically integrating Eqs.~\eqref{equation:evolution:insp}, one obtains the time evolution of $r_p(t)$ and $\Omega_p(t)$, which can then be used to construct the adiabatic waveform of the EMRI system.

\subsection{Interpolation error of Proca fluxes}
To efficiently evolve geodesic orbits within the adiabatic approximation, rapid evaluation of the gravitational and Proca energy fluxes is essential. However, direct computation of these fluxes from Eq.~\eqref{flux:total} is prohibitively slow for generating adiabatic EMRIs and their corresponding waveforms. To overcome this limitation, we employ an interpolation scheme to estimate $(\dot{E}^{\mathcal{G}}, \dot{E}^{\mathcal{P}})$ across different values of the black hole spin and Proca mass, $(a, \mu)$. Specifically, the fluxes are computed on a radial grid of 150 points uniformly distributed in the transformation $\mathcal{U} = \sqrt{r - 0.98r_{\rm ISCO}}$, spanning the range $[\mathcal{U}(r_{\rm min}), \mathcal{U}(r_{\rm max})]$, where $r_{\rm min} = r_{\rm ISCO} + 0.15M$ and $r_{\rm max} = r_{\rm min} + 10M$. The corresponding non-uniform grid in the orbital radius is obtained by inverting this mapping, $r = r(\mathcal{U})$, which we denote succinctly as $r_\mathcal{U}$. During the inspiral evolution, the orbital radius is restricted to $r_p \in [r_{\rm ISCO}, r_0]$, where $r_{\rm ISCO}$ denotes the innermost stable circular orbit of the Kerr spacetime and $r_0$ is the initial orbital radius.

For each pair of spin and Proca mass $(a,\mu)$, we compute the relativistic gravitational and Proca fluxes along the sampling grid $r_\mathcal{U}$, denoted by $\left[\dot{E}^{\mathcal{G}}(r_\mathcal{U}), \dot{E}^{\mathcal{P}}(r_\mathcal{U})\right]$. The spin parameter is uniformly sampled over $a \in [0.0001, 0.01]$, and the Proca mass over $0.0001 \leq \mu \leq 0.01$ with a step size of $\delta\mu = 0.0005$. On this $(a,\mu)$ rectangular grid of $20 \times 20$ nodes, both fluxes are evaluated at 150 radial points for each set of parameters. The resulting dataset is then used to construct the interpolation functions $\left[\dot{E}^{\mathcal{G}}_{\rm int}(r_\mathcal{U}), \dot{E}^{\mathcal{P}}_{\rm int}(r_\mathcal{U})\right]$, which enable efficient and accurate flux evaluation for arbitrary $(a,\mu)$ values within the sampled range.

\begin{figure}[htb!]
\centering
\includegraphics[width=3.2in, height=2.2in]{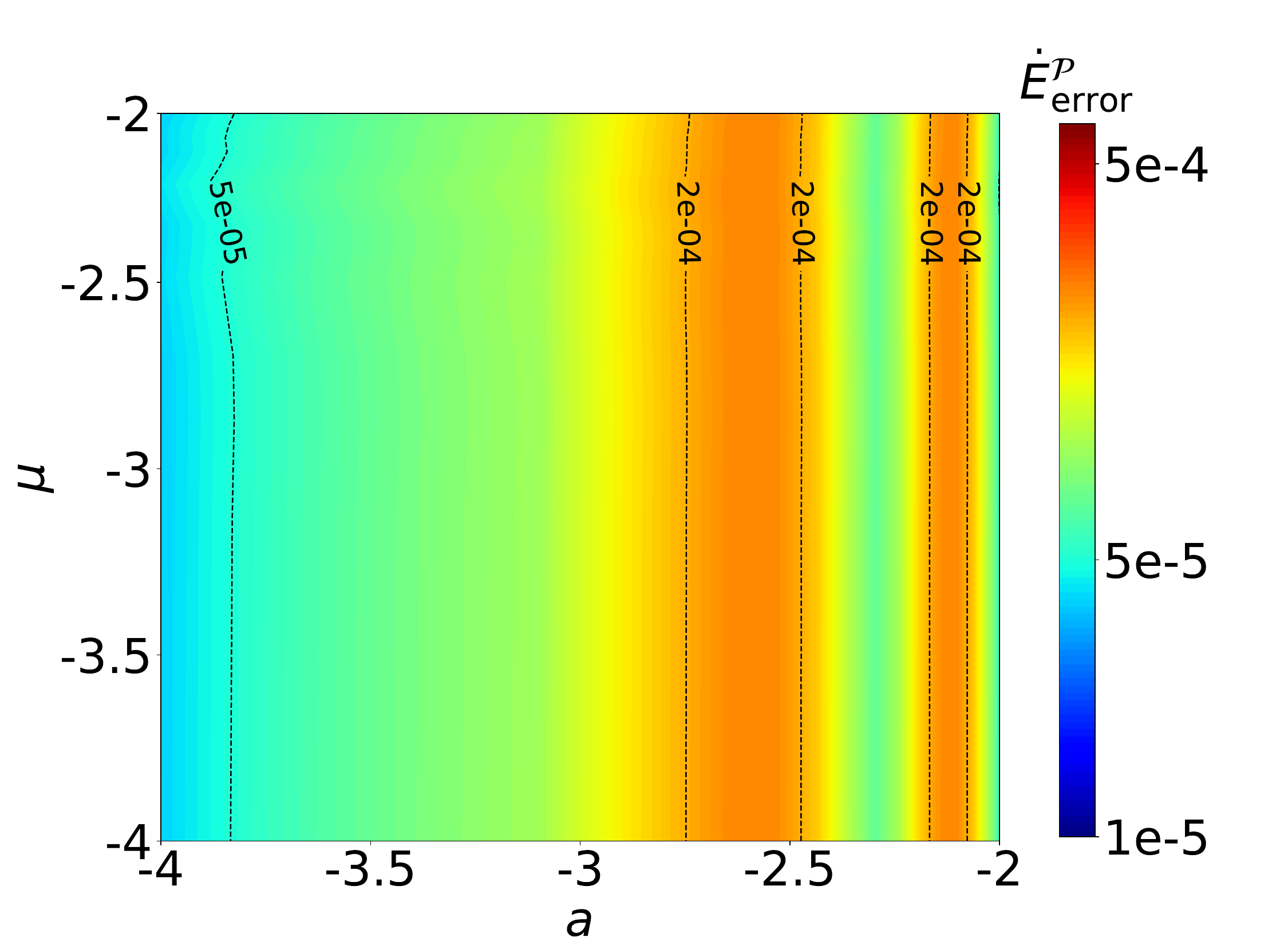}
\includegraphics[width=3.2in, height=2.2in]{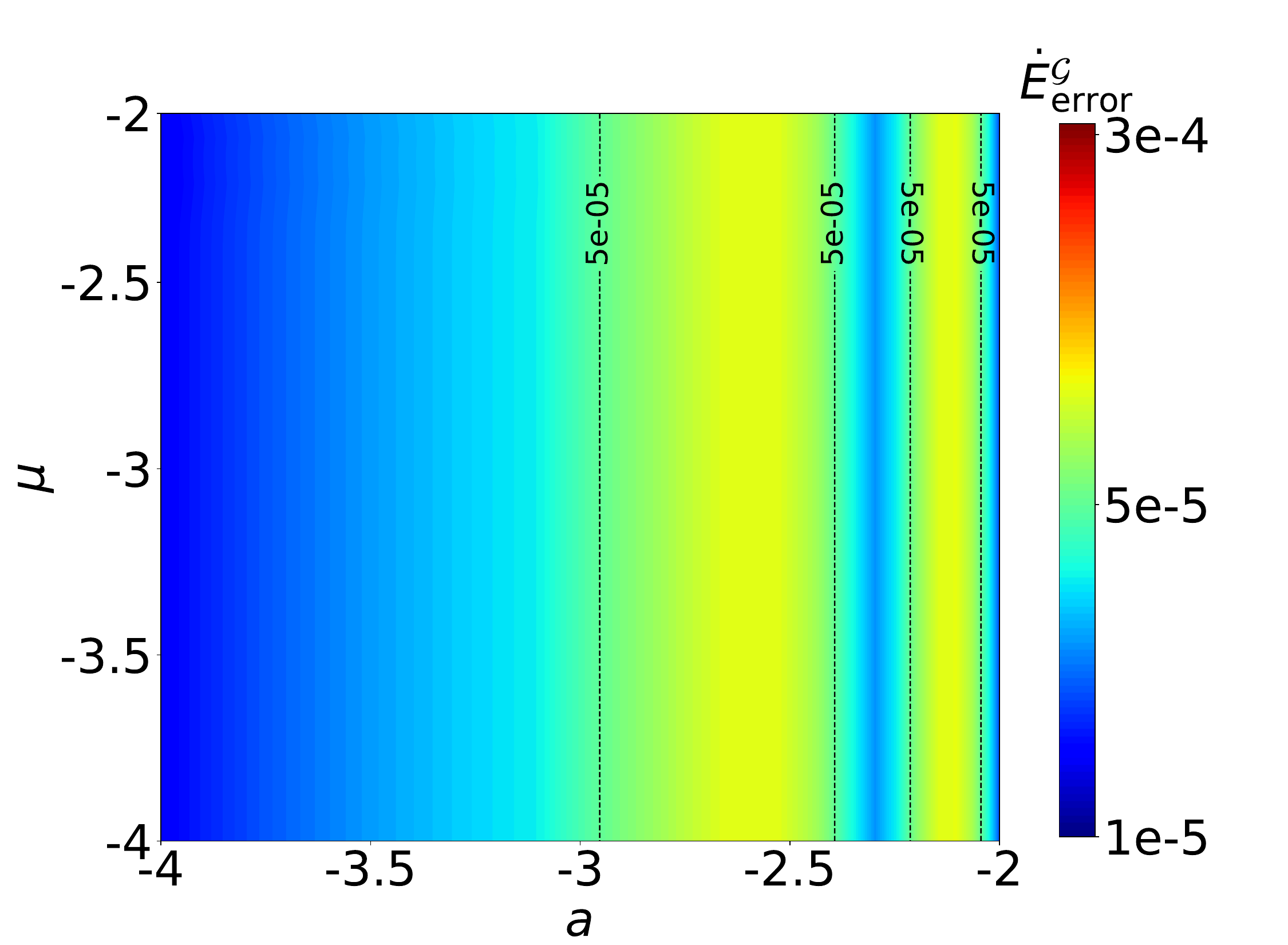}
\caption{Interpolation errors of the gravitational and Proca energy fluxes, obtained by summing over multipole indices $(l,m)$ off a rectangular grid as functions of the Proca mass and MBH spin $(\mu,a)$, assuming a fixed vector charge $q=0.1$. The symbols $\dot{E}^{\mathcal{G}}_{\rm error}$ and $\dot{E}^{\mathcal{P}}_{\rm error}$ represent the discrepancies between the fluxes computed via the interpolation method and those obtained from perturbation theory. All fluxes are expressed in units of the mass ratio $m_p^2/M^2$. Black dashed contours indicate lines of constant flux error between the two methods.} \label{fig:error:interp}
\end{figure}

To assess the robustness of the interpolation scheme, we compute the fluxes at off-grid $(a,\mu)$ values using both the interpolation model and direct perturbative evaluation from Eq.~\eqref{flux:total}. The interpolation-induced errors in the gravitational and Proca fluxes are denoted by $\dot{E}^{\mathcal{G}}{\rm error}$ and $\dot{E}^{\mathcal{P}}{\rm error}$, respectively. Figure~\ref{fig:error:interp} illustrates the error distribution across the $(a,\mu)$ parameter space. As shown, the interpolation error for the gravitational flux $\dot{E}^{\mathcal{G}}{\rm error}$ lies within the range $[10^{-5},2\times10^{-4}]$, while that for the Proca flux $\dot{E}^{\mathcal{P}}{\rm error}$ is within $[4\times10^{-5},3\times10^{-4}]$. These results demonstrate that the interpolation method maintains high fidelity across the sampled grid, indicating that the method provides a reliable and computationally efficient surrogate for flux evaluation in EMRI orbital evolution and waveform generation.

\subsection{EMRIs waveform and data analysis}
In this subsection, we outline the procedure for computing the EMRIs waveform within the quadrupole approximation, and introduce mismatch of waveform  as a diagnostic to quantify the influence of Proca-field radiation on the orbital dynamics of the EMRI, as reflected in the resulting waveform.

Once the inspiraling trajectories are obtained from Eq.~\eqref{equation:evolution:insp},
the corresponding quadrupolar EMRIs waveform can be computed from the metric perturbation
in the transverse-traceless (TT) gauge, defined as
\begin{equation}
h_{ij} = \frac{2}{d_L} \left(P_{il} P_{jm} - \frac{1}{2} P_{ij} P_{lm}\right) \ddot{I}_{lm},
\end{equation}
where \( d_L \) denotes the luminosity distance between the source and the detector, and
\( P_{ij} = \delta_{ij} - n_i n_j \) is the projection operator onto the plane orthogonal
to the wave propagation direction \( n_i \).
The second time derivative of the mass quadrupole moment, which sources the gravitational
waveform, can be determined from the stress-energy tensor of the system~\cite{Babak:2006uv}:
\begin{equation}
I_{ij} = \int d^3x\, T^{tt}(t, x^i)\, x^i x^j = m_p\, y^i(t)\, y^j(t),
\end{equation}
where \( T^{tt}(t, x^i) = m_p \delta^{(3)}(x^i - y^i(t)) \), \( x^i \) are spatial Cartesian
coordinates, and \( y^i(t) \) denotes the worldline of the secondary compact object.

The two gravitational-wave polarization amplitudes can then be expressed in terms of the
second derivatives of the mass quadrupole moment as
\begin{eqnarray}\label{amplitude}
A_n^+ &=& -\frac{1}{2}\left(\ddot{I}_{11} - \ddot{I}_{22}\right) (1 + \cos^2\iota)
         = \mathcal{A} (1 + \cos^2\iota) \cos[2\Phi(t)], \nonumber \\
A_n^\times &=& 2\ddot{I}_{12} = -2\mathcal{A}\cos\iota\,\sin[2\Phi(t)],
\end{eqnarray}
where \( \mathcal{A} = (M\omega)^{2/3} m_p / d_L \), and \( \iota \) is the inclination angle
that characterizes the orientation of the orbital plane with respect to the line of sight.
The source location and orientation are further specified by the four angular parameters
\( (\theta_S, \phi_S, \theta_L, \phi_L) \), as defined in Ref.~\cite{Barack:2003fp}.
The strain of amplitudes of GW responded by LISA can be approximately given by
\begin{equation}\label{antenna}
h_{\textup{I,II}} = \sum_n \frac{\sqrt{3}}{2} \Big[F^+_{\textup{I,II}} (t)A^+_n(t) + F^\times_{\textup{I,II}} (t)A^\times_n(t) \Big]\;,
\end{equation}
where the quantities $F^+_{\textup{I,II}}$ are the antenna pattern functions,
their detailed expressions can refer to Ref.~\cite{Cutler:1997ta}.

Prior studies indicate that the choice of waveform template does not significantly affect the measurement precision of the intrinsic parameters
$\Theta_i=(M,m_p,r_p,\Phi_{},q,\mu)$~\cite{Katz:2021yft}.
We therefore employ a hybrid construction relativistic trajectories combined with a quadrupole
waveform formula to estimate the imprint of a lighter Proca field on EMRIs signals.
A preliminary, model-agnostic assessment of the Proca mass's distinguishability is given by the
noise-weighted mismatch between waveforms with and without the Proca correction. We define
\begin{align}\label{overlap}
\mathcal{M} \equiv 1-\mathcal{O}(h_a,h_b)
= 1-\frac{(h_a|h_b)}{\sqrt{(h_a|h_a)\,(h_b|h_b)}} \,,
\end{align}
with the noise weighted inner product for LISA-like detector~\cite{Cutler:1994ys}
\begin{equation}\label{overlap2}
(a|b) \equiv 4\,\mathrm{Re}\!\int_{f_{\rm low}}^{f_{\rm high}}
\frac{\tilde a(f)\,\tilde b^{*}(f)}{S_n(f)}\,\mathrm{d}f \, .
\end{equation}
Here $S_n(f)$ is the one-sided noise power spectral density of the detector (e.g., LISA~\cite{LISA:2017pwj}),
and the integration range is $f\in[f_{\rm low},f_{\rm high}]$, with $f_{\rm low}=0.1~\mathrm{mHz}$ and
$f_{\rm high}$ set by the orbital frequency at the ISCO ($f_{\rm ISCO}$) or by the frequency reached after one year of evolution, whichever is smaller. A vanishing mismatch implies indistinguishability for detectors.

An empirical resolvability criterion for LISA-like experimental facilities is
\begin{equation}
\mathcal{M}_c \simeq \frac{1}{2\rho^2}\,,
\end{equation}
where $\rho$ is SNR of signal observed by detectors. Chosing a representative EMRIs signal with $\rho=20$~\cite{Babak:2017tow,Fan:2020zhy},
this gives $\mathcal{M}_c \simeq 1.25\times10^{-3}$; in practice we adopt $\mathcal{M}_c \sim 10^{-3}$ as a
conservative threshold for distinguishing Proca mission affected waveforms from their GR counterparts in the analysis below.

Finally, we estimate EMRIs parameters and constrain the Proca mass using the
Fisher information matrix (FIM). In gravitational-wave inference, the FIM
provides leading-order estimates of statistical uncertainties under the
assumptions of stationary, Gaussian noise, high SNR,
and weak (flat) priors~\cite{Vallisneri:2007ev}. In this regime, the posterior
is well approximated by a multivariate normal centered on the true parameters,
with the covariance given by the inverse FIM.
We consider the parameter vector
\[
\boldsymbol{\theta}=\{M,m_p,r_p,\mu,q,\Phi_{0},\theta_S,\phi_S,\theta_K,\phi_K,d_L\}\,,
\]
and define the FIM in the standard inner product as
\begin{equation}
\Gamma_{ij} \equiv
\left(\frac{\partial h}{\partial \theta_i}\,\bigg|\,
\frac{\partial h}{\partial \theta_j}\right) ,
\qquad i,j=1,\dots,13 \, .
\end{equation}
The covariance matrix is
\begin{equation}\label{eq:cov:matrix}
\Sigma_{ij} \equiv \langle \delta\theta_i \,\delta\theta_j\rangle
= \left(\Gamma^{-1}\right)_{ij} \, ,
\end{equation}
so that the 1\,$\sigma$ uncertainties follow from its diagonal,
\begin{equation}\label{sigma:fim}
\sigma_{\theta_i}=\sqrt{\Sigma_{ii}} \, .
\end{equation}
This framework yields preliminary (typically stronger than mismatch)
bounds on the Proca mass $\mu$, while the off-diagonal elements of
$\Sigma_{ij}$ quantify parameter correlations.

\section{Result}\label{result}

\begin{table*}[htbp!]
\centering
\begin{tabular}{ccccccccc}
\hline
$a$ & $ \mu $  & $r_p/M$ &  $\dot{E}^{l_{\text{max}}=10}_\infty$ &$\dot{E}^{l_{\text{max}}=11}_\infty$
& Relative  difference
&  $\dot{E}^{l_{\text{max}}=10}_H$ &$\dot{E}^{l_{\text{max}}=11}_H$ & Relative  difference \\
\hline
$\mathbf{0.0}$ & 0.01 &6.5  &$\mathbf{ 3.60015\times10^{-4}}$ &$\mathbf{3.60023\times10^{-4}}$
& $\mathbf{2.78\times10^{-5}}$  &$\mathbf{8.06724\times10^{-6}}$ &$\mathbf{8.06746\times10^{-6}}$
& $\mathbf{2.48\times10^{-5}}$
\\
&   &12 &  $\mathbf{3.00627\times10^{-5}}$  &$\mathbf{3.00653\times10^{-5}}$ & $\mathbf{6.65\times10^{-5}}$ &  $\mathbf{8.39543\times10^{-8}}$  &$\mathbf{8.39586\times10^{-8}}$ & $\mathbf{4.76\times10^{-5}}$
\\
\hline
0.001 & 0.01 &6.5  &$3.59561\times10^{-4}$ &$3.59562\times10^{-4}$  & $2.78\times10^{-6}$  &$8.03772\times10^{-6}$ &$8.03774\times10^{-6}$  & $2.48\times10^{-6}$\\
 &   &12 &  $3.00476\times10^{-5}$  &$3.00478\times10^{-5}$ & $6.65\times10^{-6}$ &  $8.20932\times10^{-8}$  &$8.20938\times10^{-8}$ & $7.31\times10^{-6}$ \\
\hline
0.01 & 0.01 &6.5  &$3.62252\times10^{-4}$ &$3.62256\times10^{-4}$  & $1.01\times10^{-4}$  &$7.85241\times10^{-6}$ &$7.85245\times10^{-6}$  & $5.09\times10^{-6}$\\
  &  &12 &  $3.01215\times10^{-5}$  &$3.01219\times10^{-5}$ & $1.65\times10^{-5} \ \%$ &  $7.07063\times10^{-8}$  &$7.07069\times10^{-8}$ & $8.48\times10^{-6}$ \\
\hline
0.1 & 0.01 &6.5  &$4.50694\times10^{-4}$   & $4.51066\times10^{-4}$& $0.82 \%$   &$7.25167\times10^{-6}$   & $7.25227 \times10^{-6}$  & $9.65\times10^{-5}$ \\
&   &12   &$3.08257\times10^{-5}$   &$3.08505\times10^{-5}$ &  $4.98\times10^{-4}$  &$2.18397\times10^{-7}$   &$2.18506\times10^{-7}$ &  $4.98\times10^{-5}$\\
\hline
\hline
$\mathbf{0.0}$ & 0.05 &6.5  &$\mathbf{3.11354\times10^{-4}}$ &$\mathbf{3.11348\times10^{-4}}$  & $<\mathbf{10^{-2} \ \%}$  &$\mathbf{7.18036\times10^{-6}}$ &$\mathbf{7.18038\times10^{-6}}$  & $<\mathbf{10^{-2} \ \%}$\\
&    &12 &  $\mathbf{2.87264\times10^{-7}}$  &$\mathbf{2.87263\times10^{-7}}$ & $\mathbf{<10^{-2} \ \%}$ &  $\mathbf{6.61246\times10^{-8}}$  &$\mathbf{6.61247\times10^{-8}}$ & $<\mathbf{10^{-2} \ \%}$ \\
    \hline
$\mathbf{0.0}$ &0.5 &6.5  &$\mathbf{3.02237\times10^{-14}}$   & $\mathbf{4.48767\times10^{-13}}$& $<\mathbf{93 \ \%}$   &$\mathbf{9.66965\times10^{-8}}$   & $\mathbf{9.66966\times10^{-8}}$& $<\mathbf{10^{-2} \ \%}$ \\
 &   &12   &$0$   &$0$ &  $<\mathbf{10^{-2} \ \%}$  &$\mathbf{6.18735\times10^{-12}}$   &$\mathbf{6.18736\times10^{-12}}$ &  $<\mathbf{10^{-2} \ \%}$\\
\hline
$0.1$ &0.05 &6.5  &$3.34517\times10^{-4}$   & $3.34518\times10^{-4}$& $3.86\times10^{-6}$   &$6.55819\times10^{-6}$   & $6.55819\times10^{-6}$   & $<10^{-8} $ \\
&   &12   &$3.30765\times10^{-7}$   &$3.30765\times10^{-7}$ &  $<10^{-10}$  &$7.38286\times10^{-8}$   &$7.38286\times10^{-8}$ &  $<10^{-10}$\\
\hline
$0.1$ &0.5 &6.5  &$3.34517\times10^{-4}$   & $3.34518\times10^{-4}$& $3.86\times10^{-6}$   &$6.55819\times10^{-6}$   & $6.55819\times10^{-6}$   & $<10^{-8} $ \\
&   &12   &$0$   &$0$ &  $-$  &$4.96779\times10^{-13}$   &$4.96779\times10^{-13}$ &  $<10^{-10}$\\
\hline
\hline
\end{tabular}
\caption{The massive vector fluxes (in units of $m_p^2/M^2$) from maximum multipolar component $l\in[10,11]$ for two orbital radiuses $r_p=6.5M$ and $r_p=12M$  are listed, which consider the different vector fields with the masses $\mu \in \{0,0.01,0.05,0.5\}$. The sixth column and ninth column are the relative difference of the vector flux for $l_{\text{max}}=10$ and $l_{\text{max}} = 11$, which correspond to the cases at the horizon and at the infinity. Note that the parameters $\mu=0$ and $\mu\neq0$ denote the cases of the massless vector field and the Proca field,  and the black and bold folds are the results for the Schwarzchild spacetime, which comes from Ref.~\cite{Zi:2024lmt}. }\label{tab:Fluxvalues}
\end{table*}

\begin{figure*}[htb!]
\centering
\includegraphics[width=3.2in, height=2.2in]{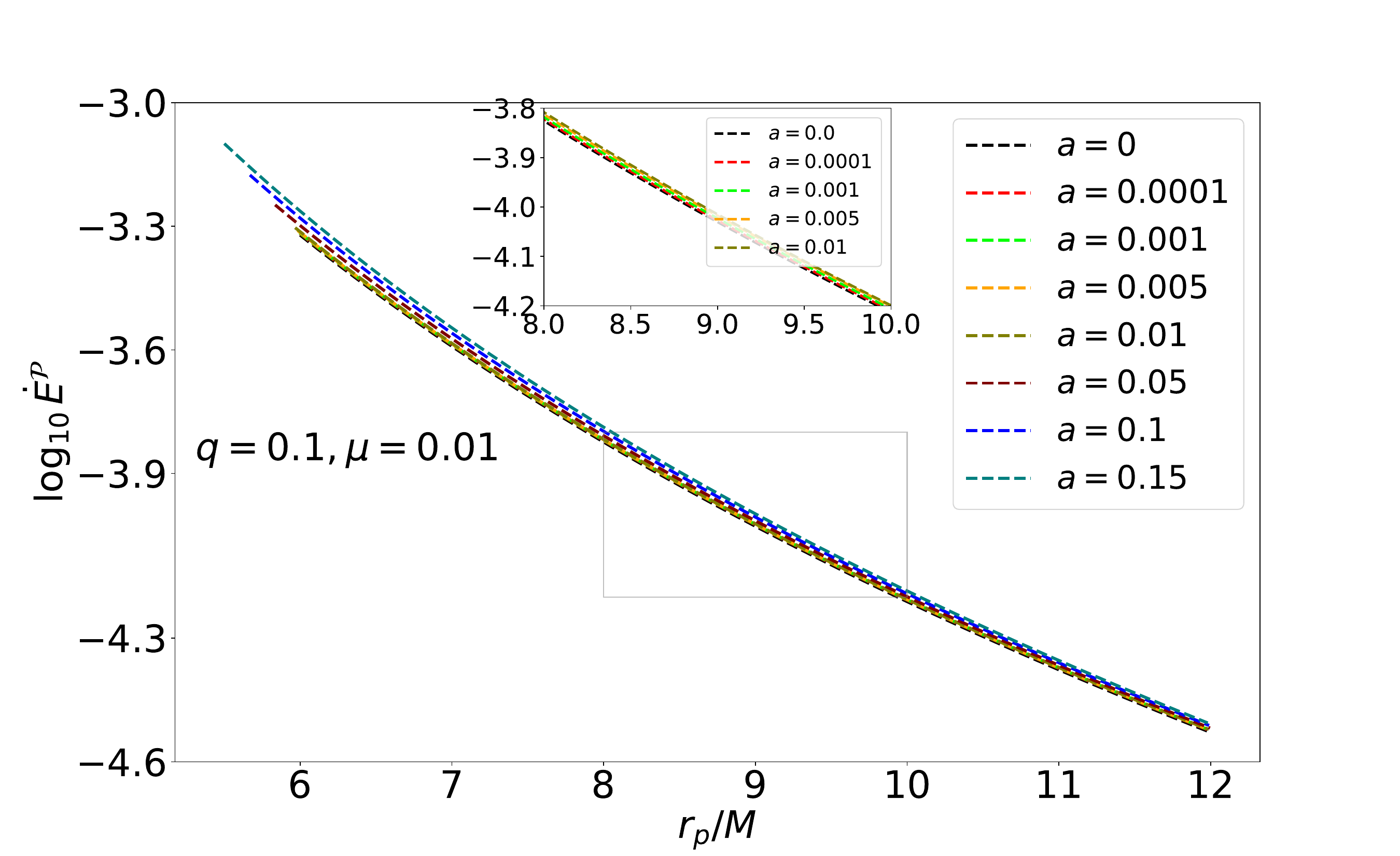}
\includegraphics[width=3.2in, height=2.2in]{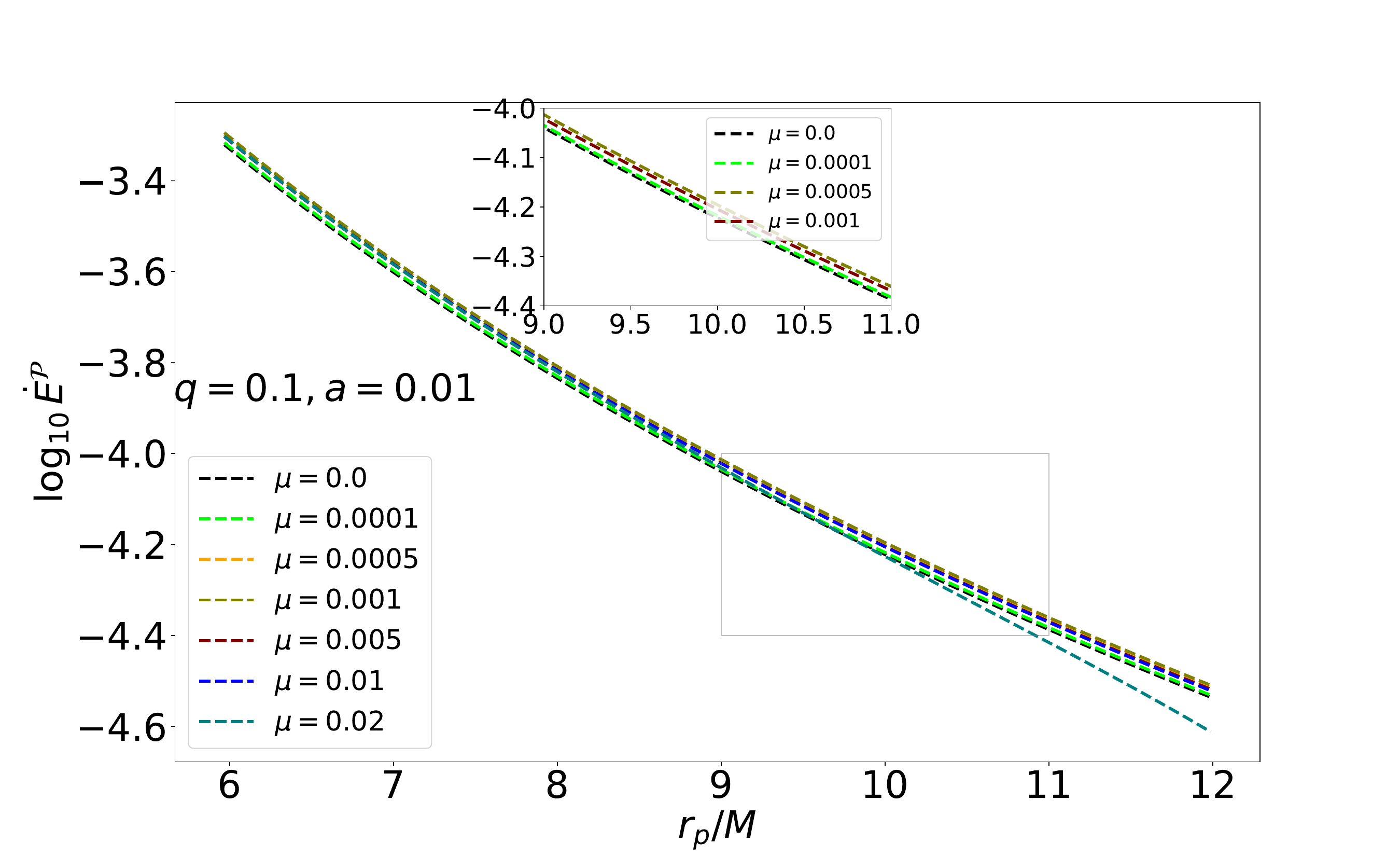}
,\includegraphics[width=3.2in, height=2.2in]{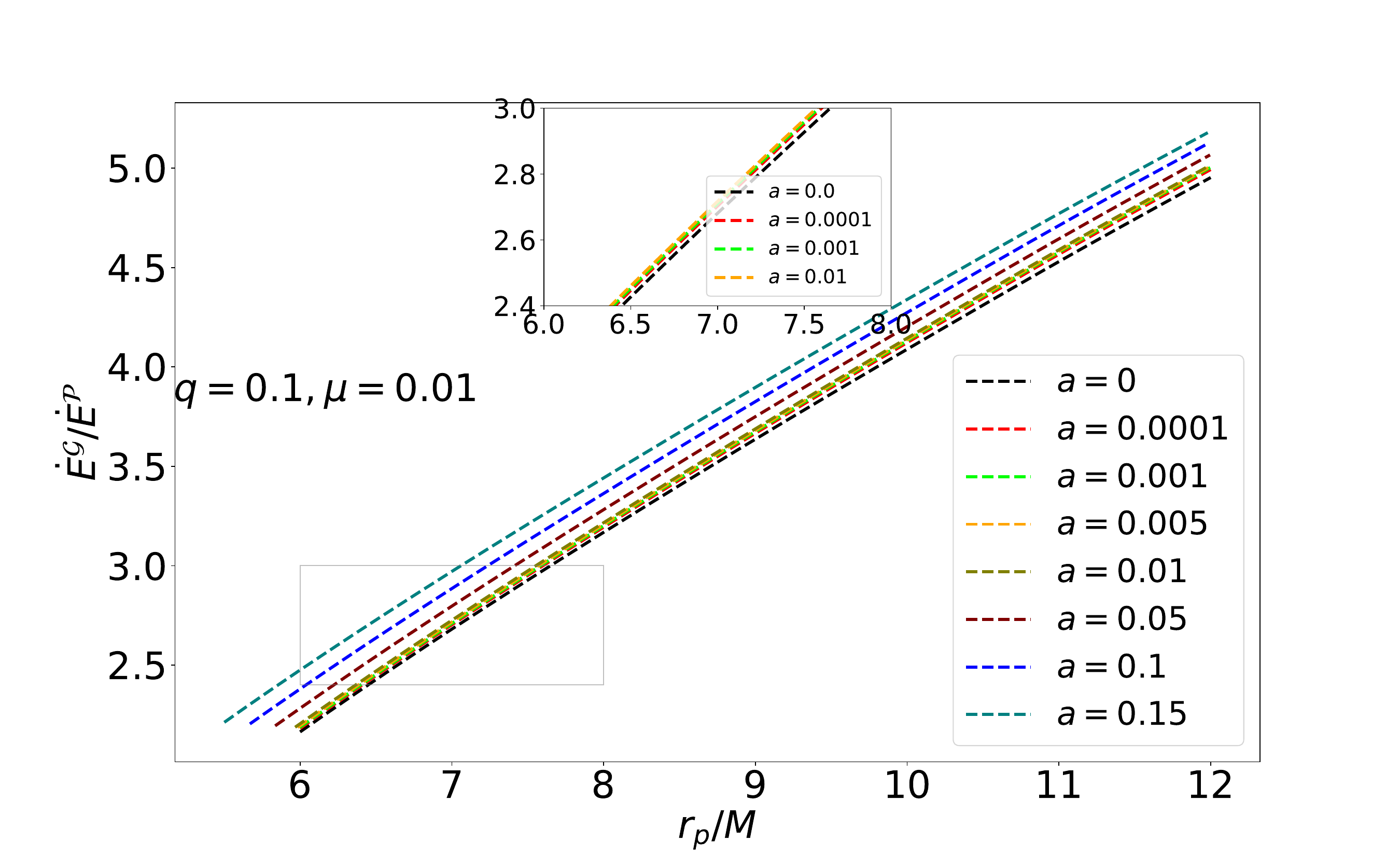}
\includegraphics[width=3.2in, height=2.2in]{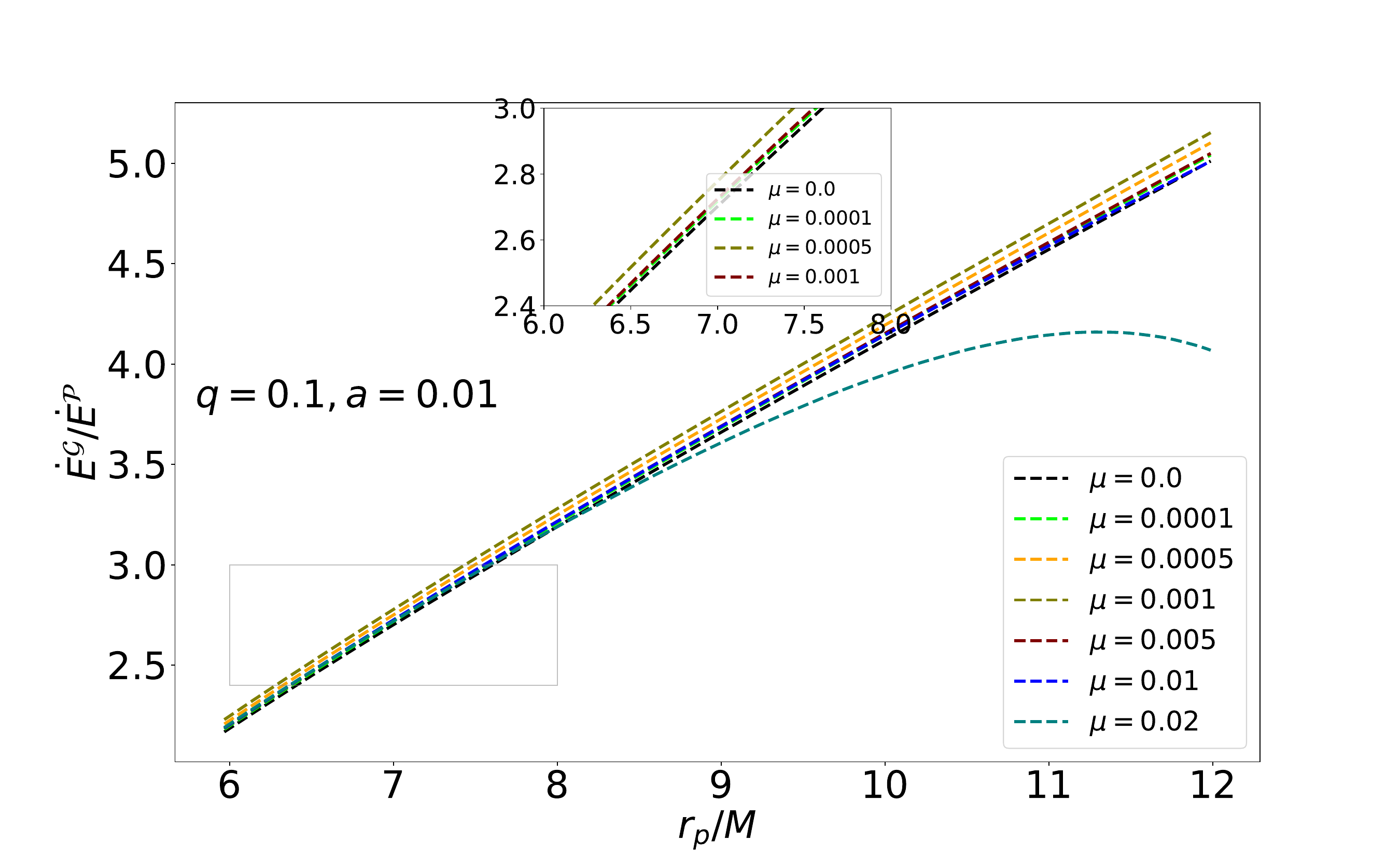}
\caption{The Proca flux and the ratio between the gravitational and Proca fluxes as a function of the orbital radius $r_p/M$ are plotted for different values of MBH spin and Proca mass,
which consider the cases of different spinning Kerr BH with $a\in \{0.0,0.0001,0.001,0.005,0.01,0.05,0.1,0.2\}$ in the left panels, and different Proca fields with a mass
$\mu \in \{0.0,0.0001,0.0005,0.001,0.005,0.01,0.02\}$ and a fixed charge $q=0.1$ in the right panels.
The subfigures are the zoom magnified pictures in the ranges $r_p/M\in[r_{\rm ISCO},12] $ of orbital radius.   Note that these fluxes have the units of $m_p^2/M^2$.}\label{energyProca}
\end{figure*}

\begin{figure*}[htb!]
\centering
\includegraphics[width=6.2in, height=3.2in]{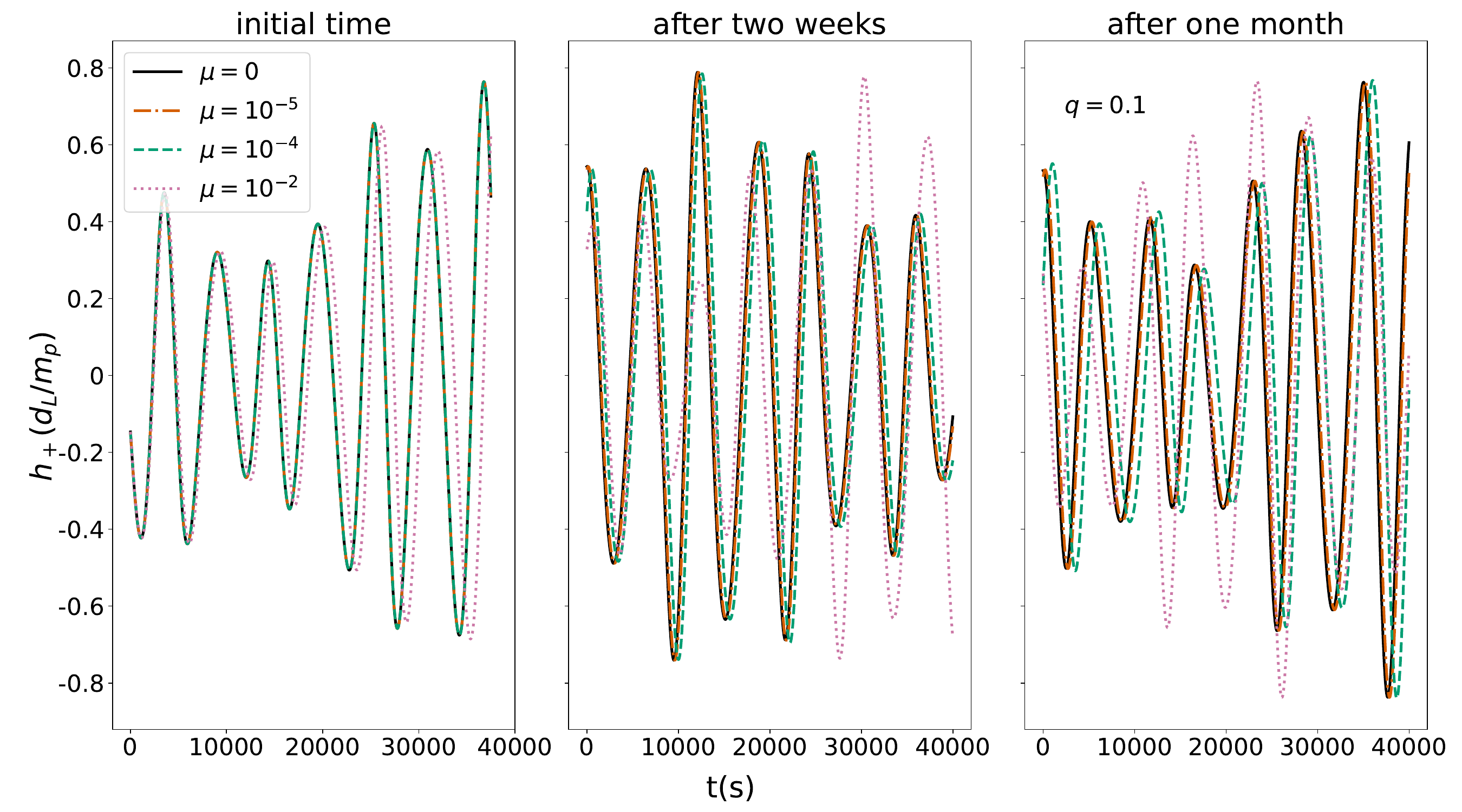}
\caption{
Comparison of the plus polarization $h_{+}(t)$ of EMRIs waveforms in the cases of GR  and  Einstein-Proca family with vector field masses $\mu = \{0,\,10^{-5},\,10^{-4},\,10^{-2}\}$ and fixed charge $q = 0.1$. The time-domain signals display three phases. Increasing $\mu$ leads to phase shifts and amplitude modulations relative to the GR case, highlighting the impact of the massive vector field on the waveform phase.
}\label{fig:wave}
\end{figure*}

\begin{figure*}[htb!]
\centering
\includegraphics[width=3.2in, height=2.2in]{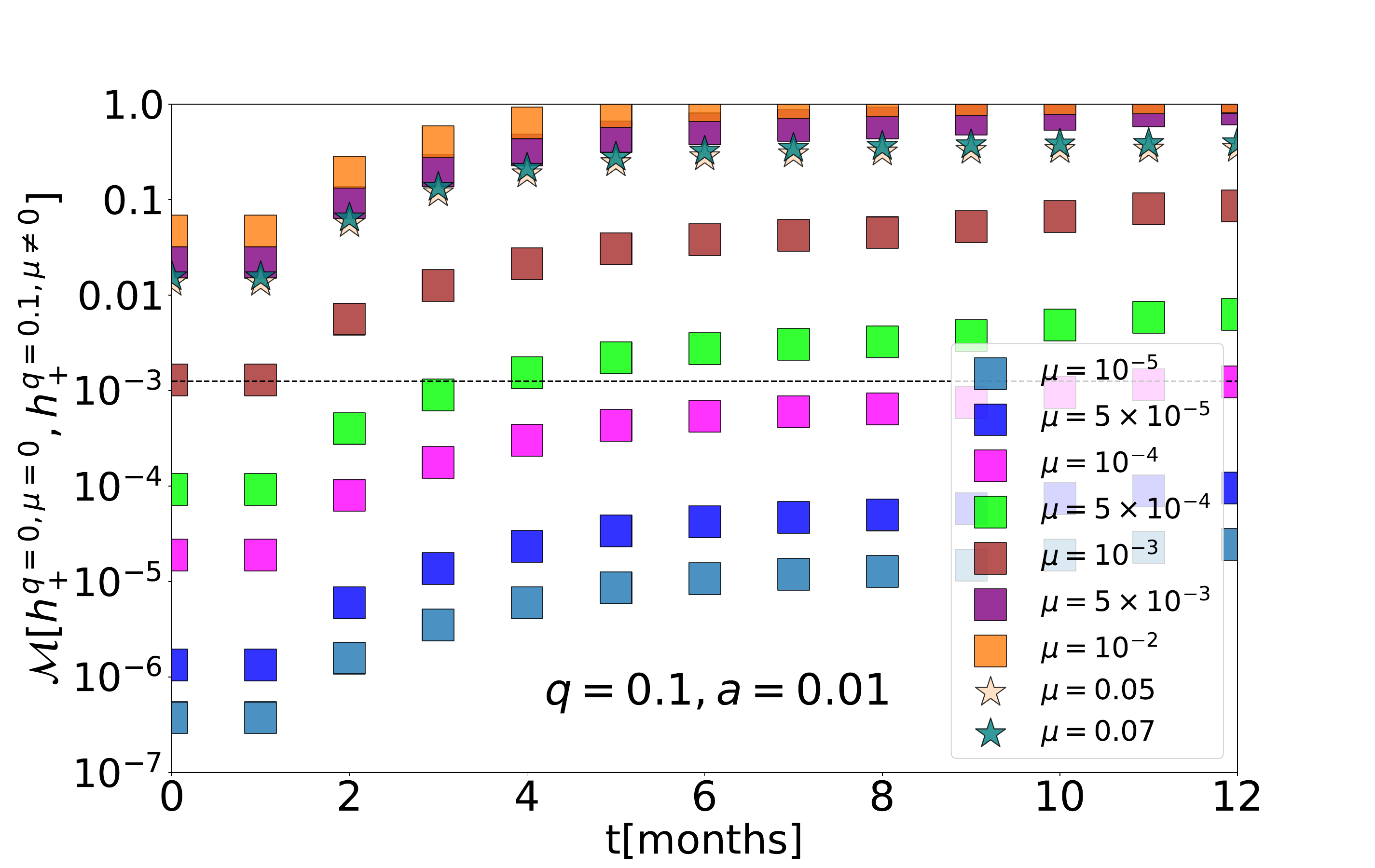}
\includegraphics[width=3.2in, height=2.2in]{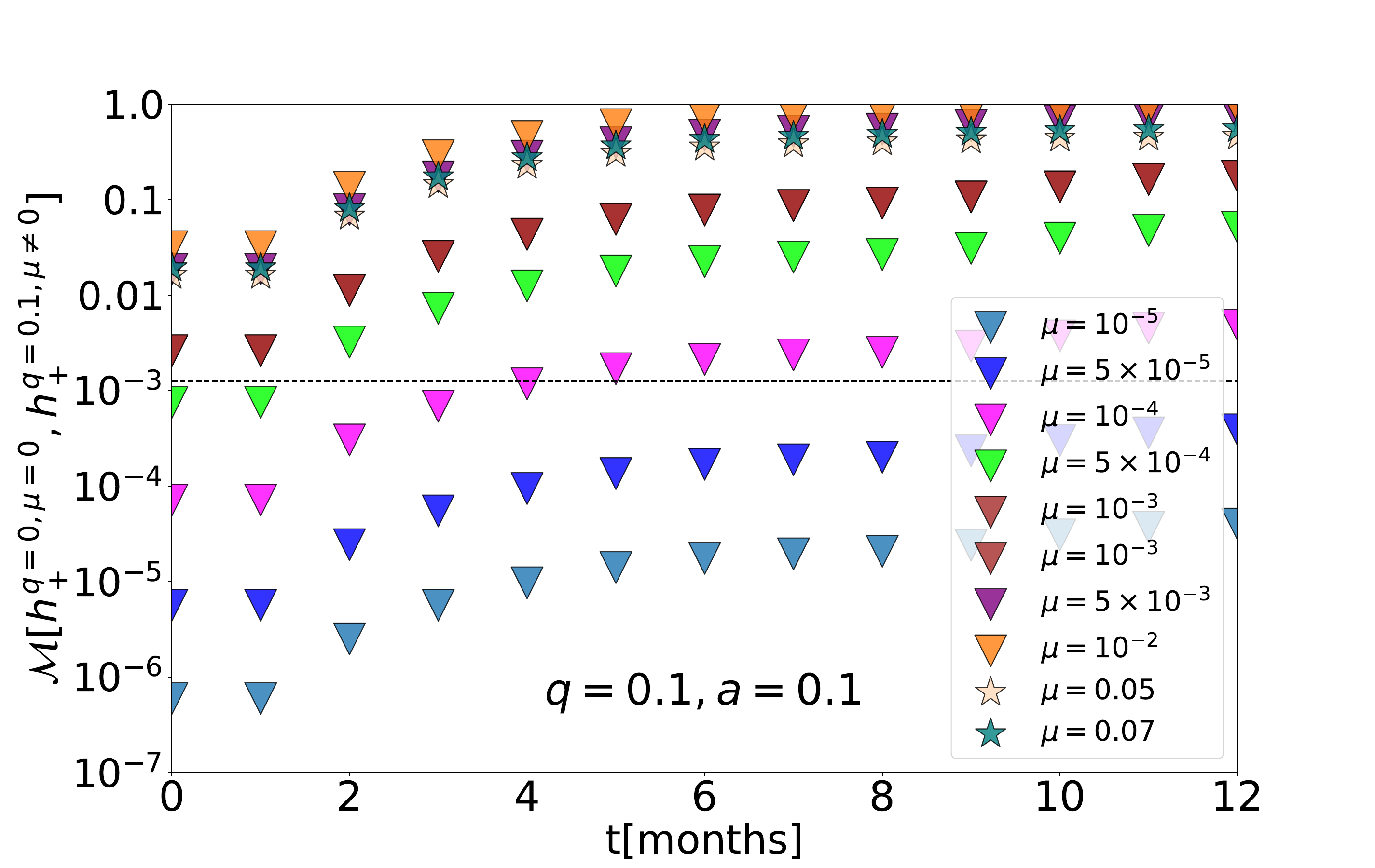}
\includegraphics[width=3.2in, height=2.2in]{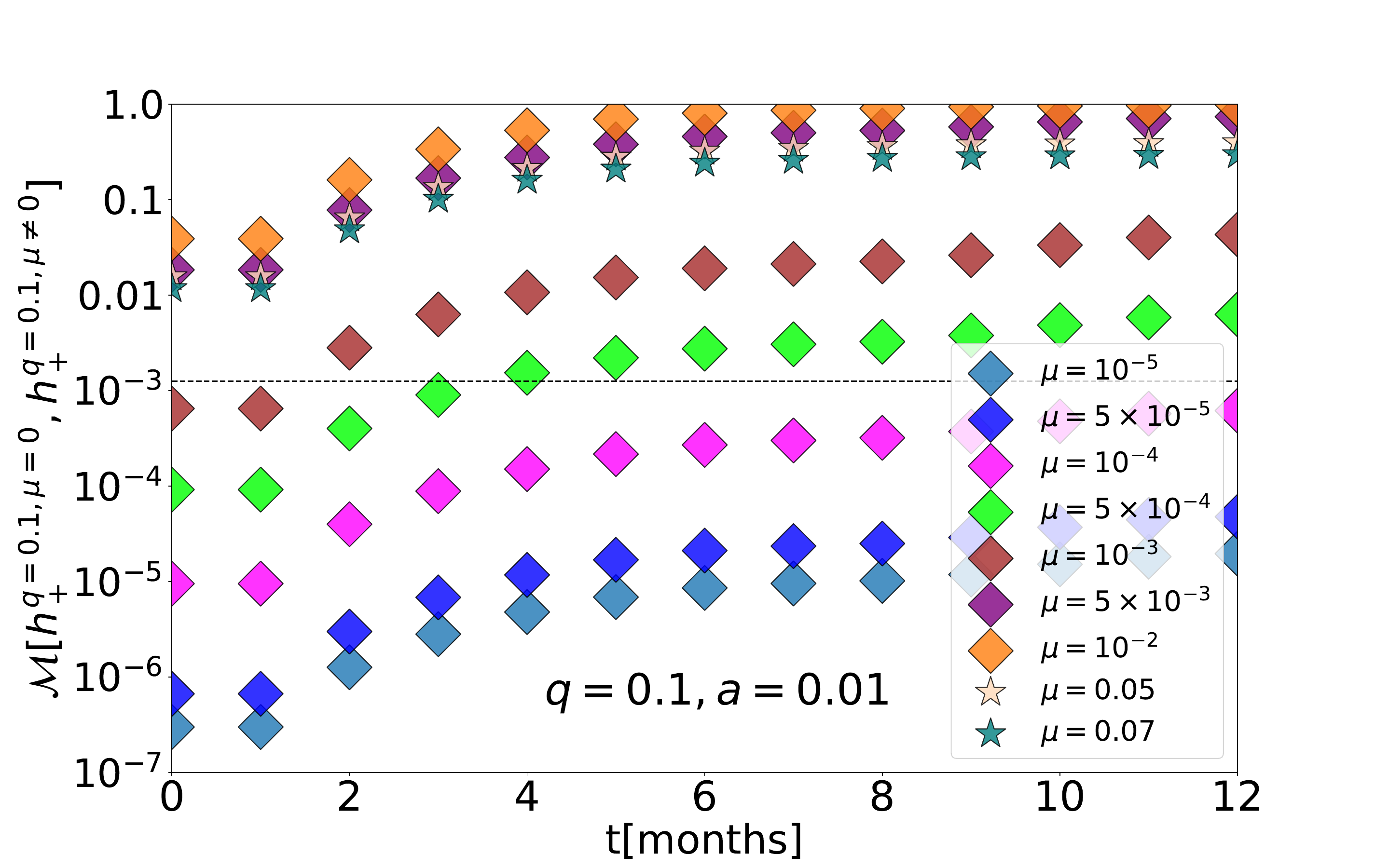}
\includegraphics[width=3.2in, height=2.2in]{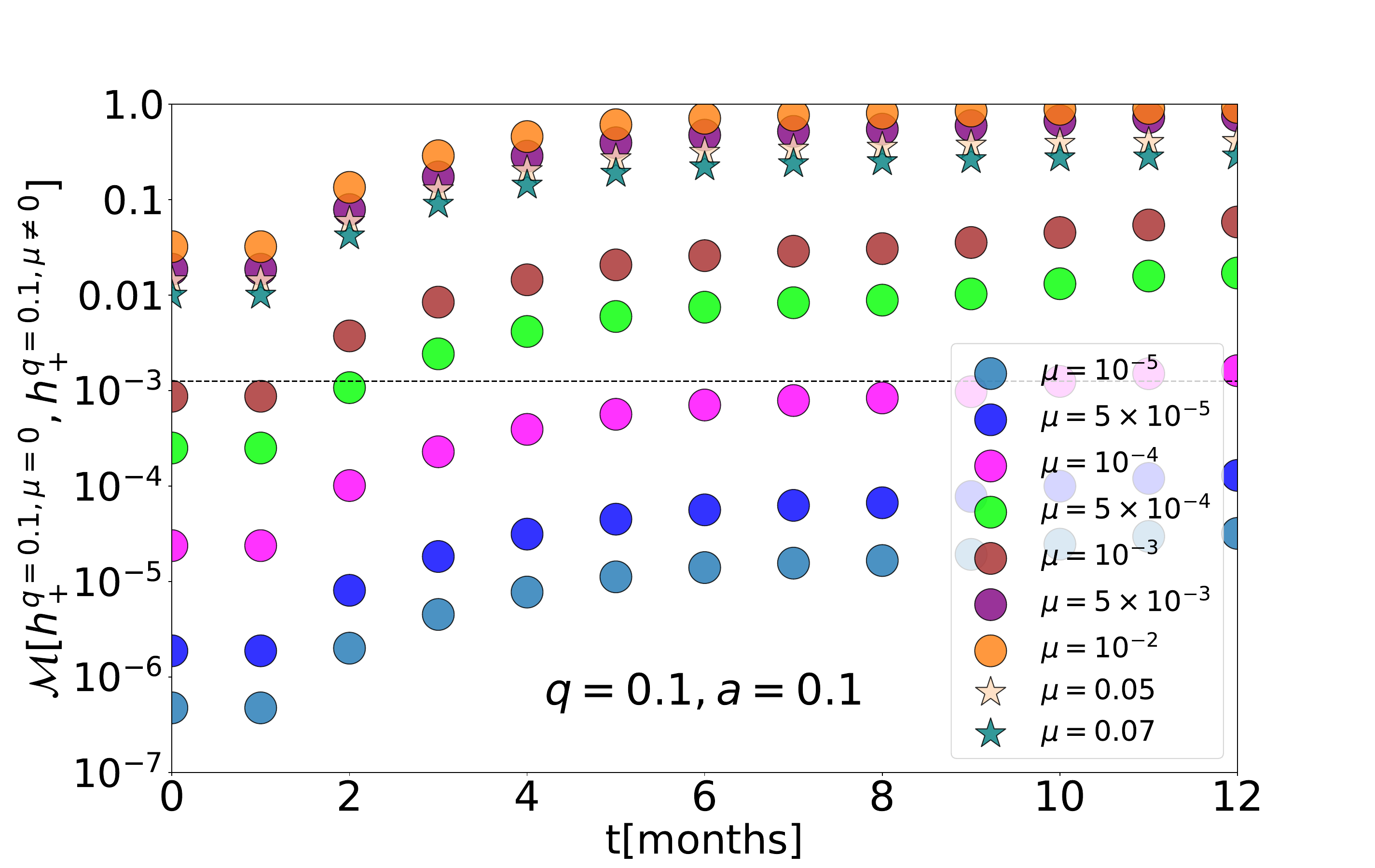}
\caption{The mismatches between the massless vector field or standard GR cases and the Proca field cases  are plotted, where the  parameters of the  vector fields carried by the smaller objects are  set as $\Big( \{ \mu=0,q=0\}, \{\mu\neq0,q\neq0 \} \Big)$ and $\Big( \{ \mu=0,q=0.1\}, \{\mu\neq0,q=0.1 \} \Big)$, and the other parameters are set as $\{r_p=12M, m_p=10 \Msun, M=10^6 \Msun\}$.
    Note that the horizontal black lines denote to the threshold value of mismatch for the top and bottom panels.
}\label{fig:mismatch}
\end{figure*}

In this section, we first compare Proca and gravitational fluxes by listing some numerical results by
setting different orbital parameters in Sec.~\ref{proca:flux}, and analyse the distinguishability of EMRIs signal with the modification of Proca fluxes in Sec.~\ref{waveform:diff}. Then we present the constraint on Prcoca mass using the statistical method of FIM.

In this paper, we fix the directional parameters of EMRI system to $\theta_S=\pi/3$, $\phi_S=\pi/2$, $\theta_K=\pi/4$ and $\phi_K=\pi/4$.
The initial orbital separation $r_0$ is chosen such that the system undergoes one year of adiabatic evolution before the final plunge, defined by $r_{\text{end}} = r_{\text{ISCO}} + 0.1M$.
We consider an EMRI with component masses $m_p = 10M_\odot$ and $M = 10^6M_\odot$, vector charge $q=0.1$ and explore various values of the Proca field mass $\mu$.
The luminosity distance $d_L$ is treated as a free parameter, allowing SNR of signal to be varied.

\subsection{Proca flux}\label{proca:flux}
In this subsection, we investigate the Proca fluxes emitted by EMRIs system. Our analysis shows that the fluxes vanish at infinity when the orbital frequency is below the Proca mass, $\omega_m < \mu$. This suppression closely parallels the behavior observed for massive scalar fields \cite{Berti:2012bp,Barsanti:2022vvl}. In contrast, the horizon flux remains nonzero for all orbital radii, continuously influencing the orbital evolution of EMRIs throughout the inspiral phase.

The total Proca flux is obtained by summing over all multipolar components $(l,m)$ with $1 \leq l \leq l_{\rm max}$ and $1 \leq m \leq l$, where we adopt $l_{\rm max} = 10$. The relative differences among massive vector fluxes for various configurations are summarized in Table~\ref{tab:Fluxvalues}. For radiation at infinity, the difference between consecutive modes $(l_{\rm max}, l_{\rm max}+1)$ remains below $0.01\%$ when the Proca mass satisfies $0 \leq \mu \leq 0.05$. Near the horizon, the flux differences between successive modes stay below $10^{-2}\%$ across the broader range $0 \leq \mu \leq 0.5$. When the field mass increases to $\mu = 0.5$, the total flux at infinity reduces to $\lesssim 93\%$ of its massless value, implying that higher-order multipoles ($l > 10$) may contribute appreciably as the compact object approaches the central massive black hole. Furthermore, we also find that the Proca flux  would increase slightly when the spinning parameter of central MBH is bigger for the orbital radius $r_p=6.5M$.

Figure~\ref{energyProca} presents the energy flux ratios between the Proca and gravitational channels, together with the logarithmic Proca flux, $\log_{10}(\dot{E}^{\mathcal{P}})$, as a function of the orbital radius $r_p$. The analysis spans black-hole spins $a \in \{0, 0.001, 0.01, 0.1\}$, vector masses $\mu \in \{0, 0.0001, 0.0005, 0.001, 0.005, 0.01, 0.02\}$, and includes multipole contributions up to $l_{\rm max}=10$. The top panels correspond to two EMRIs configurations: (left panel) varying the spin of MBH for fixed vector mass $\mu = 0.01$, and (right panel) varying $\mu$ for fixed spin $a = 0.01$. The bottom panels display the corresponding ratios of Proca to gravitational flux.
For a given vector charge and mass $(q=0.1,\,\mu=0.01)$, the Proca flux increases mildly with the MBH spin, indicating enhanced energy extraction in more rapidly rotating backgrounds. Conversely, for fixed spin $a = 0.01$, the flux decreases as the vector mass grows, consistent with the suppression of radiation by the Proca mass term. This trend is corroborated by the flux ratios shown in the lower panels: the Proca flux rises gradually as the secondary inspirals toward the ISCO, remaining comparable in magnitude to the massless limit for slowly rotating Kerr black holes, but exhibits a sharp decline for heavier vector fields.

Overall, while the Proca mass suppresses the total flux and thereby weakens its imprint on the EMRI waveform, the deviation from the general-relativistic case remains significant for sufficiently massive vector fields. Moreover, for a fixed Proca mass, EMRIs around more rapidly spinning MBHs exhibit stronger departures from GR due to their enhanced vector flux emission.

\begin{table*}[htbp!]
\centering
\begin{tabular}{cc|cccccccccccc}
\hline
\hline
$\mu$ & $a$ & $\sigma_M/M$ & $\sigma_{m_p}/m_p$ &$\sigma_{a}$   &$\sigma_{r_p}$  & $\sigma_{\mu}/\mu$  & $\sigma_{q}$
& $\sigma_{\theta_S}/\theta_S$ & $\sigma_{\phi_S}/\phi_S$
 & $\sigma_{\theta_K}/\theta_K$   & $\sigma_{\phi_K}/\phi_K$
& $\sigma_{\Phi{_{\phi,0}}}/\Phi_{\phi,0}$  & $\sigma_{d_L}$
\\
\hline
0.01  & 0.0  &$1.90\text{e-5}$  &$3.27\text{e-4}$  &$-$ &$1.26\text{e-4}$  & $774\%$
& $3.86\text{e-3}$  &$0.86$  & $6.36\text{e-1}$
&$7.57\text{e-1}$ &$4.57$  &$8.78\text{e-1}$    &$8.24$
\\
  &0.001   &$7.62\text{e-6}$  &$2.67\text{e-4}$ &$5.45\text{e-5}$   &$3.47\text{e-4}$  & $438\%$
&$3.43\text{e-3}$  &$0.58$  &$4.54\text{e-1}$
&$5.78\text{e-1}$ &$1.48$
&$7.48\text{e-1}$    &$6.78$
\\
 &0.1  &$3.23\text{e-6}$  &$1.30\text{e-5}$  &$4.67\text{e-5}$   &$2.45\text{e-4}$
& $156\%$   &$2.89\text{e-3}$  &$1.15\text{e-1}$
&$2.07\text{e-1}$  &$1.01\text{e-1}$ &$1.01\text{e-1}$
 &$1.48\text{e-1}$    &$3.22$
\\
\hline
\hline
0.02  & 0.0  &$2.50\text{e-5}$  &$5.26\text{e-4}$  &$-$  &$1.57\text{e-4}$  & $29.65\%$
& $6.43\text{e-3}$  &$0.76$  & $6.54\text{e-1}$
&$7.68\text{e-1}$ &$3.87$
&$8.15\text{e-1}$   &$8.45$
\\
&0.001   &$2.48\text{e-5}$  &$2.67\text{e-4}$   &$2.87\text{e-5}$   &$3.15\text{e-4}$    & $10.24\%$
&$2.24\text{e-3}$  &$2.31\text{e-1}$  &$4.01\text{e-1}$  &$1.62\text{e-1}$  &$1.44\text{e-1}$
&$7.35\text{e-2}$   &$5.24$
\\
&0.1  &$2.15\text{e-5}$  &$1.85\text{e-4}$   &$1.65\text{e-5}$  &$2.24\text{e-4}$
& $7.65\%$   &$7.27\text{e-4}$  &$1.72\text{e-1}$
&$8.62\text{e-2}$  &$8.67\text{e-2}$  &$6.23\text{e-2}$
 &$2.11\text{e-2}$   &$5.34$  \\
\hline
\hline
\end{tabular}
\caption{Measurement errors for EMRIs with masses $(M=10^{6}M_\odot,m_p= 10M_\odot)$,
vector charge and vector $(q=0.05,\mu\in\{0.0,0.01,0.02\})$, initial orbital separation $(r_p=10.0)$ and
spin $a\in\{0,0.001,0.1\}$, initial orbital phases $(\Phi_{\phi,0}=1.0)$ and
the directional angles related to source $(\theta_{S,K}=\phi_{S,K}=1.0)$ are listed, where the secondary object lasts the spiraling of two years and the luminosity distance $(d_L)$ is adjusted to set SNR of EMRIs signal as $150$. The transverse line below $\sigma_{e_0}$ means the circular EMRIs.
}\label{tab:fim:error}
\end{table*}

\begin{figure*}[htb!]
\centering
\includegraphics[width=0.87\paperwidth]{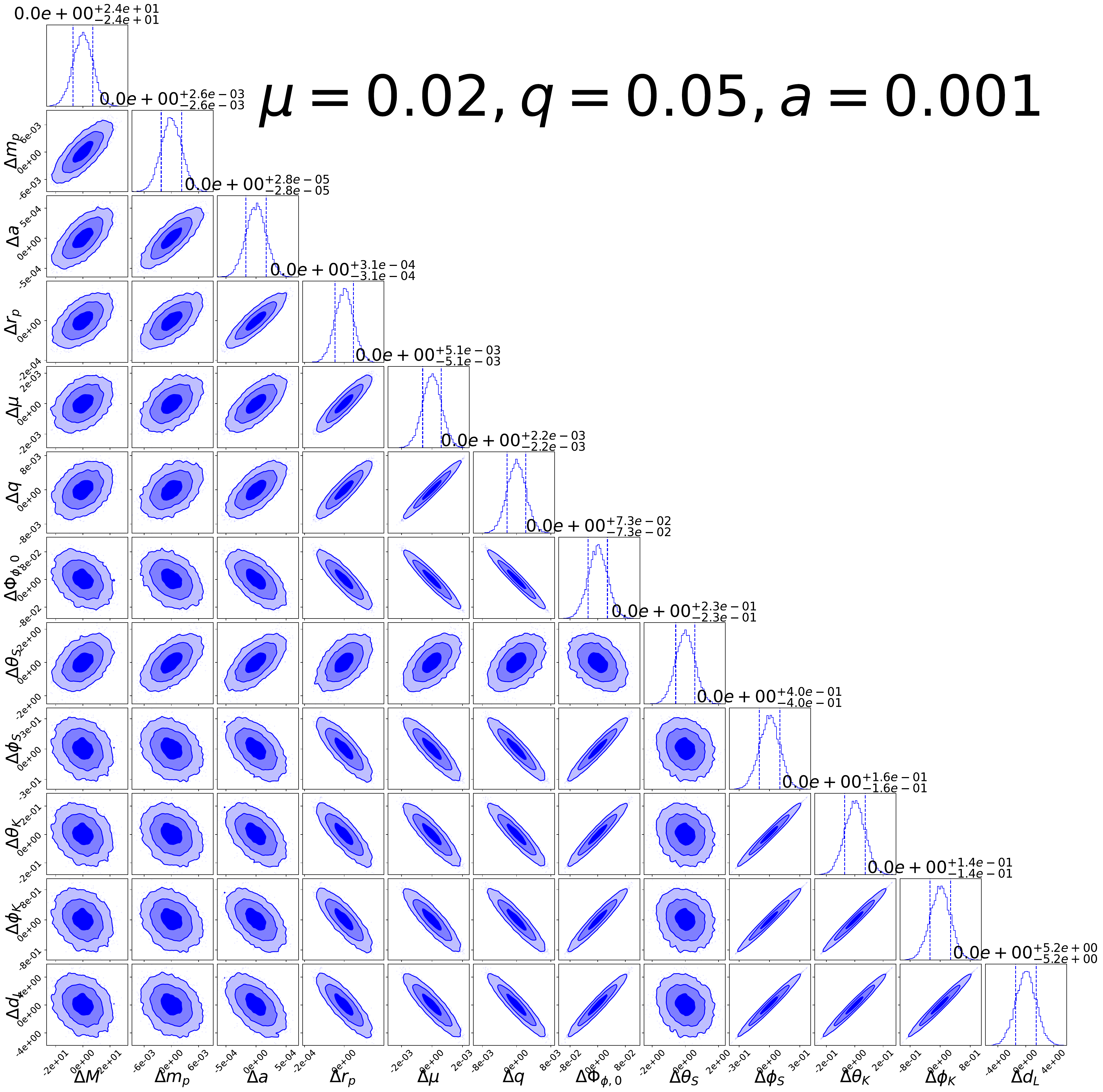}
\caption{Corner plot of the posterior probability distributions for EMRI source parameters: the component masses, MBH spin, initial orbital radius, initial orbital azimuthal phase, vector  mass and charge are assumed as follows $(M = 10^6 M_\odot,\, m_p = 10 M_\odot,\, a = 0.001,\, r_p = 12M,\, \Phi_{\phi,0}=1.0,\,\mu = 0.02,\, q = 0.05)$. Other parameters are fixed as described in Sec.~\ref{result}. The distributions are inferred from a two-year LISA observation. Vertical dashed lines indicate the $1\sigma$ credible intervals for each parameter, while the contours represent the 68\%, 95\%, and 99\% confidence levels.
}\label{fig:cornerplot:a0d001}
\end{figure*}

\begin{figure*}[htb!]
\centering
\includegraphics[width=0.87\paperwidth]{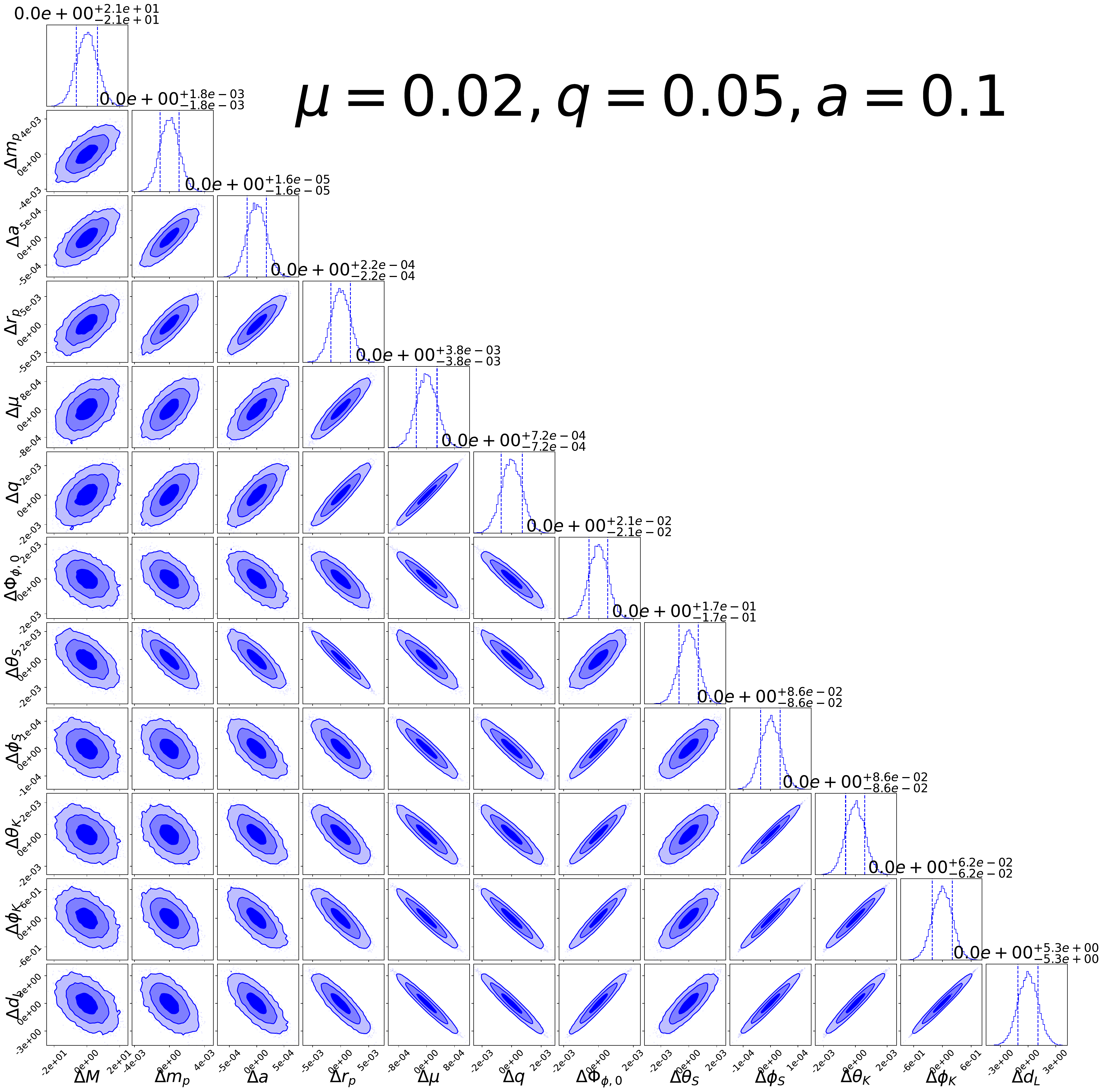}
\caption{Corner plot of the marginalized posterior probability distributions for the EMRI source parameters. The component masses, MBH spin, initial orbital radius, initial azimuthal phase, vector mass, and vector charge are fixed to $(M = 10^6 M_\odot, m_p = 10 M_\odot, a = 0.1, r_p = 12M, \Phi_{\phi,0} = 1.0, \mu = 0.02, q = 0.05)$, while the remaining parameters are chosen as in Sec.~\ref{result}. The posteriors are inferred from a two-year LISA observation. Vertical dashed lines mark the $1\sigma$ credible intervals for each parameter, and the contours denote the $68\%$, $95\%$, and $99\%$ confidence regions.
} \label{fig:cornerplot:a0d1}
\end{figure*}

\begin{figure*}[htb!]
\centering
\includegraphics[width=3.17in, height=2.3in]{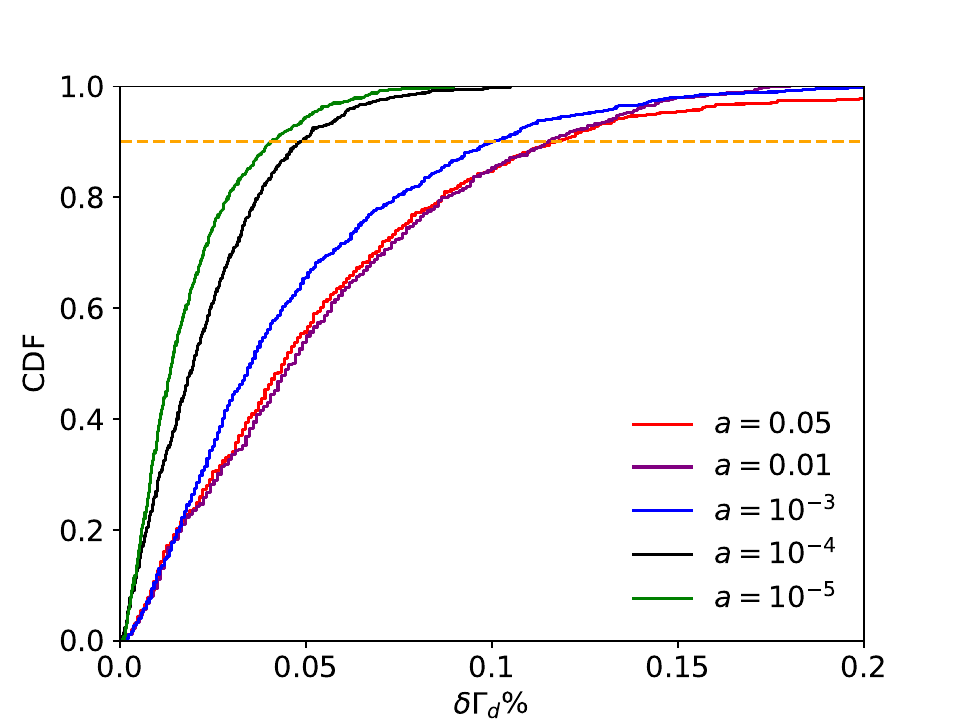}
\includegraphics[width=3.17in, height=2.3in]{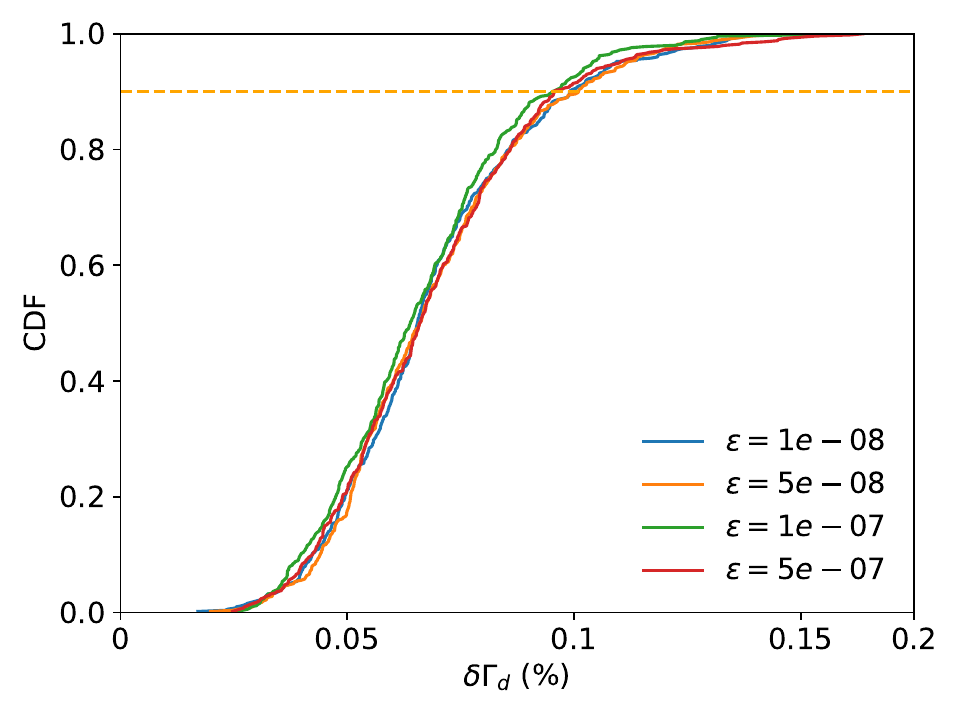}
\caption{
Cumulative distribution of the maximum relative error between the FIM and its perturbed counterpart, where each element of the deviation matrix follows a uniform distribution $u \in [-10^{-3},10^{-3}]$. Different colored curves correspond to several numerical derivative spacings $\epsilon\in[10^{-8},5\times10^{-8},10^{-7},5\times10^{-7}]$ used in the computation of the FIM in the right panel. The left panel shows the cases of different MBH spin $a\in[10^{-5},10^{-4},10^{-3},0.01,0.05]$.
Other parameters setting in the computing two panels keep same with Fig.~\ref{fig:cornerplot:a0d1}.
} \label{Fig:CDF}
\end{figure*}

\subsection{Waveform and its differentiability}\label{waveform:diff}
In this subsection, we compare EMRIs waveforms predicted by GR and the Einstein-Proca theory, and investigate the impact of the Proca mass on EMRIs signal morphology. The plus polarization of the waveform, rescaled by $d_L/m_p$, is shown in Fig.~\ref{fig:wave} for different vector masses $\mu = \{0,\,10^{-5},\,10^{-4},\,10^{-2}\}$ with a fixed vector charge $q = 0.1$.

As illustrated in Fig.~\ref{fig:wave}, the phases of the waveforms remain nearly identical during the initial $\sim 10^3$ seconds of evolution, indicating negligible influence from the Proca mass at early times. For larger vector masses, such as $\mu = 10^{-2}$, a noticeable phase deviation becomes apparent after approximately $1500$ seconds. Over longer timescales, the accumulated phase difference grows significantly: for $\mu = 10^{-4}$, a clear phase shift arises after about two weeks of evolution. After one month, all massive Proca cases exhibit distinct phase offsets relative to the massless case, demonstrating that the presence of a finite vector mass introduces cumulative dephasing effects in EMRIs signal.

To assess the influence of the Proca field mass on the GW signals from EMRIs around a rotating Kerr black holes, we evaluate two types of waveform mismatches. The first quantifies the deviation between the standard GR and Einstein-Proca cases, while the second compares the massless and massive vector radiation scenarios.
Fig.~\ref{fig:mismatch} displays four mismatches for various combinations of the MBH spin and vector mass, with the vector charge fixed at $q = 0.1$. The upper panels show the mismatches between the GR and Einstein-Proca theory predicting waveforms for spin parameters $a = 0.01$ (left) and $a = 0.1$ (right), while the lower panels illustrate the corresponding results for the massless and massive vector cases. As shown from Fig.~\ref{fig:mismatch}, the magnitude of mismatch grows with both the Proca mass and the MBH spin is increasing, indicating that even a light vector field can induce measurable phase shifts over long observation timescales. The horizontal dashed lines mark the distinction threshold $\mathcal{M}_c$ by LISA, above which the deviation from GR becomes distinguishable. For a fixed vector charge, the mismatch generally increases with the Proca mass; however, when the mass becomes sufficiently large, the distinguishability weakens slightly due to suppression of the Proca flux, consistent with the behavior presented in Table~\ref{tab:Fluxvalues} and Fig.~\ref{energyProca}.

The upper left panel of Fig.~\ref{fig:mismatch}, corresponding to a slowly rotating MBH with $a = 0.01$, compares EMRIs systems characterized by two vector parameters $(q, \mu) = (0, 0)$ and $(0.1, \mu \neq 0)$. The results show that a six-month observation of LISA could detect Proca masses as small as $\mu_{\mathrm{min}} \sim 5 \times 10^{-4}$ for $q = 0.1$. Increasing the spin to $a = 0.1$, as shown in the upper right panel, enhances the mismatch, suggesting that moderate MBH rotation improves the detectability of Proca-induced effects. The lower panels further illustrate the differences between the massless and massive vector radiation for $a = 0.001$ and $a = 0.1$, revealing that LISA can distinguish Proca field effects for masses as low as $\mu_{\mathrm{min}} \sim 4 \times 10^{-4}$. Overall, while a more massive Proca field produces a larger mismatch, its contribution to the orbital evolution becomes negligible for $\mu \gtrsim 0.05$, as the corresponding radiation flux is strongly suppressed beyond a critical mass threshold of $\mu \sim 0.5$, as shown in Table~\ref{tab:Fluxvalues}. Consequently, the mismatch between GR and Einstein-Proca waveforms first increases with the field mass, reaches a peak at intermediate values, and gradually declines toward zero as the Proca flux near the horizon becomes negligible for large $\mu$.

\subsection{Constraint on Proca mass}\label{constraint}
In this subsection, we constrain the Proca field mass using EMRIs signals from a slowly rotating Kerr MBH. Since full Bayesian inference with Markov Chain Monte Carlo techniques \cite{Karnesis:2023ras} demands waveform generation on millisecond timescale \cite{Katz:2021yft} and thus requires substantial computational resources, we adopt the FIM approach to obtain preliminary bounds on the impact of Proca radiation on EMRI waveforms.

Table~\ref{tab:fim:error} summarizes the parameter estimation uncertainties derived from the FIM analysis, assuming a SNR of 150 achieved by adjusting the luminosity distance $d_L$.
For comparison with the circular EMRIs around a Schwarzschild black hole, we adopt the same vector charge $q=0.05$ as in Ref.~\cite{Zi:2024lmt}. The first and fourth rows correspond to signals from a Schwarzschild MBH, while the remaining entries are computed under identical source parameters to those in Ref.~\cite{Zi:2024lmt}. The inclusion of spin significantly improves the measurement precision: the relative uncertainty in the Proca mass decreases from $774\%$ to $156\%$ for $\mu=0.01$. For a more massive vector field ($\mu=0.02$), the measurement error of the Proca mass can reach $6.40\%$ ($\sim8.53\times10^{-18}\,\mathrm{eV}$), representing a substantial enhancement in parameter precision. This improvement reflects the stronger deviation of the massive vector radiation from the massless case, as discussed in Ref.~\cite{Zi:2024lmt}.
Moreover, a secondary object endowed with Proca hair can inspiral deeper into the strong-field region of a rotating MBH than the Schwarzschild case, thereby encoding richer information about the system parameters. One can find that intrinsic quantities such as the component masses, vector charge, and orbital elements show modest improvement with LISA observations, the extrinsic parameters (sky position and distance) remain comparatively unaffected. These results highlight the crucial role of MBH spin in detecting the signatures of Proca-induced deviations from GR.

Finally, we perform a correlation analysis among source's parameters using the off-diagonal elements of the covariance matrix defined in Eq.~\eqref{sigma:fim}. As illustrated in Figs.~\ref{fig:cornerplot:a0d001} and \ref{fig:cornerplot:a0d1}, distinct correlation patterns emerge across the parameter space. In particular, the positive correlations between the primary mass $M$ and other parameters become slightly stronger as the MBH spin increases from $a = 0.001$ to $a = 0.1$. Similar trends are observed for the secondary mass $m_p$, which exhibits enhanced correlations with several parameters, including $a$, $r_p$, $q$, $\mu$, $d_L$, and $(\theta_{S,K}, \phi_{S,K})$.
The Proca mass $\mu$ shows positive correlations with intrinsic parameters $(M, m_p, r_p, q, a)$ for both spin values considered, $a = (0.001, 0.1)$. In contrast, $\mu$ exhibits strong negative correlations with the extrinsic parameters $(\theta_{S,K}, \phi_{S,K}, d_L)$ that describe the source's sky localization and distance. A similar correlation pattern is found for the vector charge $q$, indicating consistent coupling behavior between the Proca field properties and the source's intrinsic parameters. These correlation analysis suggest that precise measurements of the intrinsic parameters are crucial for placing tighter bounds on the Proca mass and for breaking potential degeneracies in EMRI parameter estimation.

When computing the measurement error of source's parameters from FIM, it is necessary to
obtain the covariance matrix defined by inversion of FIM. To assess the numerical stability of covariance matrix, we use the following method developed by several works~\cite{Speri:2021psr,Maselli:2021men,Zi:2022hcc}.
First, we take a perturbation matrix $\mathbf{U}$ with same dimension of FIM $(\mathbf{\Gamma})$,
in which the elements of $\mathbf{U}$  are randomly allocated from a uniform distribution $u \in [-10^{-3},10^{-3}]$. Second, we compute the inverse of perturbed matrix $(\mathbf{U}+\mathbf{\Gamma})$,
and evaluate the maximum relative deviation for the perturbed and standard covariance matrices, defined by
\begin{equation}
\delta \mathbf{\Gamma}_d \equiv \max \left( \frac{(\mathbf{U}+\mathbf{\Gamma})^{-1}-\mathbf{\Gamma}^{-1}}{\mathbf{\Gamma}^{-1}} \right).
\end{equation}
Figure~\ref{Fig:CDF} displays the cumulative distribution of
$\delta \mathbf{\Gamma}_d$ under two conditions: varying massive black hole (MBH)
spin values, $a \in \{10^{-5}, 10^{-4}, 10^{-3}, 0.01, 0.05\}$ (left panel), and
different numerical derivative step sizes used in evaluating the waveform's
partial derivative with respect to the Proca mass $\mu = 0.02$ (right panel).
In both cases, the results demonstrate numerical robustness more than 90\% of
the samples exhibit relative deviations below $0.15\%$, confirming that the maximum
relative errors should be tolerable in computation of inverse of FIM.

\section{Conclusion}\label{conclusion}

In this work, we have investigated the effect of Proca radiation on EMRIs signals within the Einstein-Proca framework. Using perturbation theory, we computed both the Proca and gravitational energy fluxes for a slowly rotating Kerr MBH, where the Proca perturbation equation was expanded to linear order in the spin parameter $a$. Based on these fluxes, we constructed an interpolation scheme on a two-dimensional rectangular grid of $(a, \mu)$ to generate adiabatic trajectories parameterized by the vector charge and mass.

Our mismatch analysis shows that, for a slowly rotating MBH with $a=0.01$ and a fixed vector charge $q=0.1$, LISA would be capable of distinguishing the deviations from the GR waveform when the Proca mass reaches $\mu \sim 5 \times 10^{-4}$ ($\sim 6.67 \times 10^{-20}\,\rm eV$). When the spin increases to $a=0.1$, the threshold detectable Proca mass decreases to $\mu \sim 10^{-4}$ ($\sim 1.33 \times 10^{-20}\,\rm eV$). Comparisons between the Proca and massless vector radiation cases show slightly smaller mismatches, with the Proca mass $\mu \sim 10^{-4}$ only marginally discernible for $a=0.1$. These results indicate that the EMRI waveform and fluxes modified by the Proca hair of the secondary are highly sensitive to the MBH spin, highlighting the need for accurate modeling within a fully relativistic framework.

We further evaluated the capability of LISA to detect ultra-light vector radiation using the FIM method. For a Proca mass $\mu=0.02$, the relative measurement error of the vector mass decreases from $29.65\%$ to $7.65\%$ when the background spacetime changes from a Schwarzschild to a slowly rotating Kerr ($a=0.1$) metric. This improvement arises because circular EMRIs around rotating MBHs can spiral more efficiently into the strong-field region than those around non-rotating MBHs. Correlation analysis based on the off-diagonal elements of the covariance matrix reveals positive correlations between the Proca mass $\mu$ and intrinsic parameters $(M, m_p, r_p, q, a)$ for both spin values $a = (0.001, 0.1)$, while $\mu$ exhibits strong negative correlations with the extrinsic parameters $(\theta_{S,K}, \phi_{S,K}, d_L)$. Moreover, numerical stability analysis of the covariance matrix, obtained by introducing small perturbations in the FIM, shows that over $90\%$ of the relative error distribution remains below $0.15\%$, confirming the robustness of our parameter estimation.

Overall, our results demonstrate that the MBH spin plays a key role in constraining the mass of vector fields beyond GR through EMRIs observations. Future work will focus on reducing interpolation-induced errors in flux evaluation, for instance, by employing Chebyshev polynomial interpolation as suggested in Ref.~\cite{Khalvati:2025znb}. Additionally, applying a Bayesian statistical framework~\cite{Katz:2021yft,Speri:2024qak} to relativistic flux and waveform modeling will be essential for constraining the Proca mass with higher precision. Since MBHs in galactic centers are generally fast rotating, it will also be important to account for floating orbits driven by superradiant mechanisms~\cite{Cardoso:2011xi,Barsanti:2022vvl}, which could introduce distinct features in EMRIs signals in the presence of ultra-light vector fields.

\section{Acknowledgments}
This work was funded by the National Natural Science Foundation of China with Grants No. 12405059, NO. 12205104, NO. 12165013, NO. 12347140 and No. 12375049, and Key Program of the Natural Science Foundation of Jiangxi Province under Grant No.20232ACB201008.

\appendix
\section{Energy flux formulas for the massive vector field}\label{appendix1}

Starting from the Lagrangian in Eq.~\eqref{Lagrangian}, one can derive the effective canonical stress-energy tensor $T^C_{\mu\nu}$ for vector perturbations using Noether's theorem, which reduces to the following expression:
\begin{equation}
4\pi T^C_{\mu\nu} = \frac{\partial \mathcal{L}}{\partial A^{\alpha,\mu}} A^{\alpha}_{~~,\nu} - g_{\mu\nu} \mathcal{L},
\end{equation}
where $A^{\alpha,\mu} \equiv g^{\mu\nu} \frac{\partial A^\alpha}{\partial x^\nu}$ and $A^\alpha_{~~,\nu} \equiv \frac{\partial A^\alpha}{\partial x^\nu}$.
Using the field equations, this expression can be simplified as
\begin{equation}
4\pi T^C_{\mu\nu} = F_{\mu\lambda}\partial_\nu A^\lambda - g_{\mu\nu} \mathcal{L}.
\end{equation}
A symmetric stress-energy tensor can then be constructed following the standard procedure:
\begin{eqnarray}
4\pi T_{\mu\nu} &=& F_{\mu\lambda}\partial_\nu A^\lambda - F_{\mu}^{~~\lambda}\partial_\lambda A_\nu - A_\nu \partial_{\lambda}F_{\mu}^{~~\lambda} - g_{\mu\nu}\mathcal{L} \nonumber \\
&+& \frac{\partial}{\partial x^{\lambda}}(F_{\mu}^{~~\lambda} A_\nu) \nonumber \\
&=& g^{\alpha\beta}F_{\mu\alpha}F_{\nu\beta} + \mu^2 A_{\mu}A_{\nu} - g_{\mu\nu}\mathcal{L}.
\end{eqnarray}

The corresponding vector energy fluxes are obtained by integrating the components $T_{ti}$, which describe the energy crossing a unit surface normal to the $x^i$ direction per unit time \cite{Martel:2003jj, Zhu:2018tzi}:
\begin{eqnarray}\label{dedt}
\left\langle\frac{dE}{dt}\right\rangle_{H,\infty} &=& \lim_{r \to r_h, \infty} \int T_{ti} n^i r^2 d\Omega \nonumber \\
&=& - \lim_{r \to r_h, \infty} \epsilon \int T_{tr} r^2 f\, d\Omega,
\end{eqnarray}
where $\epsilon = 1$ corresponds to the flux at spatial infinity and $\epsilon = -1$ corresponds to the flux near the horizon.

\subsection{Energy flux at infinity}

The asymptotic form of the vector potential at infinity is given by
\begin{eqnarray}\label{inf1}
A_t &=& \frac{\mathcal{A}^{1+}}{r} e^{-i\omega_m t} e^{i \sqrt{\omega_m^2 - \mu^2} r_\ast} r^{\frac{i\mu^2}{\sqrt{\omega_m^2 - \mu^2}}} Y^{lm}, \nonumber \\
A_r &=& \frac{\mathcal{A}^{2+}}{r f} e^{-i\omega_m t} e^{i \sqrt{\omega_m^2 - \mu^2} r_\ast} r^{\frac{i\mu^2}{\sqrt{\omega_m^2 - \mu^2}}} Y^{lm}, \nonumber \\
A_\theta &=& \frac{1}{l(l+1)}\left[\mathcal{A}^{3+} Y^{lm}_{,\theta} + \frac{1}{\sin\theta} \mathcal{A}^{4+} Y^{lm}_{,\phi}\right] \nonumber \\
&\times& e^{-i\omega_m t} e^{i \sqrt{\omega_m^2 - \mu^2} r_\ast} r^{\frac{i\mu^2}{\sqrt{\omega_m^2 - \mu^2}}}, \nonumber \\
A_\phi &=& \frac{1}{l(l+1)}\left[\mathcal{A}^{3+} Y^{lm}_{,\phi} - \sin\theta\, \mathcal{A}^{4+} Y^{lm}_{,\theta}\right] \nonumber \\
&\times& e^{-i\omega_m t} e^{i \sqrt{\omega_m^2 - \mu^2} r_\ast} r^{\frac{i\mu^2}{\sqrt{\omega_m^2 - \mu^2}}}.
\end{eqnarray}

Applying the Lorenz condition yields the following relation between $\mathcal{A}^{1+}$ and $\mathcal{A}^{2+}$:
\begin{equation}\label{inf2}
\omega_m \mathcal{A}^{1+} + \sqrt{\omega_m^2 - \mu^2}\, \mathcal{A}^{2+} = 0.
\end{equation}
Substituting Eqs.~\eqref{inf1} and \eqref{inf2} into Eq.~\eqref{dedt}, the energy flux at infinity is obtained as
\begin{eqnarray}
\lim_{r \to \infty} \left\langle \frac{dE}{dt} \right\rangle &=& \sum_{l=1}^\infty \sum_{m=1}^l \frac{\omega_m \sqrt{\omega_m^2 - \mu^2} (|\mathcal{A}^{4+}_{lm}|^2 + |\mathcal{A}^{3+}_{lm}|^2)}{2\pi l(l+1)} \nonumber \\
&+& \frac{\mu^2}{2\pi} \frac{\sqrt{\omega_m^2 - \mu^2}}{\omega_m} |\mathcal{A}^{2+}_{lm}|^2.
\end{eqnarray}

\subsection{Energy flux near the horizon}

Near the event horizon, the vector potential takes the form
\begin{eqnarray}\label{hor1}
A_t &=& \frac{\mathcal{A}^{1-}}{r} e^{-i\omega_m t} e^{-i\omega_m r_\ast} Y^{lm}, \nonumber \\
A_r &=& \frac{\mathcal{A}^{2-}}{r f} e^{-i\omega_m t} e^{-i\omega_m r_\ast} Y^{lm}, \nonumber \\
A_\theta &=& \frac{1}{l(l+1)}\left[\mathcal{A}^{3-} Y^{lm}_{,\theta} + \frac{1}{\sin\theta} \mathcal{A}^{4-} Y^{lm}_{,\phi}\right] e^{-i\omega_m t} e^{-i\omega_m r_\ast}, \nonumber \\
A_\phi &=& \frac{1}{l(l+1)}\left[\mathcal{A}^{3-} Y^{lm}_{,\phi} - \sin\theta\, \mathcal{A}^{4-} Y^{lm}_{,\theta}\right] e^{-i\omega_m t} e^{-i\omega_m r_\ast},
\end{eqnarray}
where $Y^{lm}_{,\theta} \equiv \frac{\partial Y^{lm}}{\partial \theta}$ and $Y^{lm}_{,\phi} \equiv \frac{\partial Y^{lm}}{\partial \phi}$.
The Lorenz condition implies the following relation between $\mathcal{A}^{1-}$ and $\mathcal{A}^{2-}$:
\begin{equation}\label{hor2}
\mathcal{A}^{1-} - \mathcal{A}^{2-} = 0.
\end{equation}
Substituting Eqs.~\eqref{hor1} and \eqref{hor2} into Eq.~\eqref{dedt}, the energy flux at the horizon becomes
\begin{equation}
\begin{split}
\lim_{r \to r_h} \left\langle \frac{dE}{dt} \right\rangle =& \sum_{l=1}^\infty \sum_{m=1}^l
\frac{\omega_m^2 |\mathcal{A}^{4-}_{lm}|^2}{2\pi l(l+1)}
+ \frac{\omega_m^2 |\mathcal{A}^{3-}_{lm}|^2}{2\pi l(l+1)} \\
&+ \frac{l(l+1) + 4\mu^2}{8\pi} |\mathcal{A}^{2-}_{lm}|^2 \\
&- \frac{i\omega_m}{4\pi} \left(\mathcal{A}^{2-}_{lm} \mathcal{A}^{3- *}_{lm} - \mathcal{A}^{3-}_{lm} \mathcal{A}^{2- *}_{lm}\right).
\end{split}
\end{equation}

\end{document}